\documentclass[final,5p,times,twocolumn]{elsarticle}

\usepackage{amssymb}
\usepackage{gensymb}
\usepackage{amsmath}
\usepackage[dvipsnames]{xcolor}
\usepackage{lineno}
\usepackage{float}
\usepackage{enumitem}
\usepackage{listings}
\definecolor{codegreen}{rgb}{0,0.6,0}
\definecolor{codegray}{rgb}{0.5,0.5,0.5}
\definecolor{codepurple}{rgb}{0.58,0,0.82}
\definecolor{backcolour}{rgb}{0.95,0.95,0.92}
\lstdefinestyle{mystyle}{
    backgroundcolor=\color{backcolour},
    commentstyle=\color{codegreen},
    keywordstyle=\color{blue},
    numberstyle=\tiny\color{codegray},
    stringstyle=\color{codepurple},
    basicstyle=\ttfamily\footnotesize,
    breakatwhitespace=false,         
    breaklines=true,                 
    captionpos=b,                    
    keepspaces=true,                 
    numbersep=5pt,                  
    showspaces=false,                
    showstringspaces=false,
    showtabs=false,                  
    tabsize=2
}
\usepackage{parskip}
\usepackage{multicol}
\usepackage{makecell}
\usepackage{tabularray}
\usepackage{hyperref}
\hypersetup{
    colorlinks=true,
    linkcolor=blue,
    filecolor=magenta,      
    urlcolor=cyan,
    }

\begin{document}

\begin{frontmatter}

\title{Scheduling follow-up observations of energetic transients with the neutrino target scheduler (NuTS)}

\author[a]{Tobias Heibges\corref{cor1}}
\ead{theibges@mines.edu}
\author[b,c]{Claire Gu\'epin\corref{cor1}}
\ead{claire.guepin@lupm.in2p3.fr}
\author[d]{Luke Kupari}
\author[a]{Hannah Wistrand}
\author[a]{Randy Lahm}
\author[f]{Johannes Eser}
\author[d]{Mary Hall Reno}
\author[e]{Tonia M. Venters}
\author[a]{Lawrence Wiencke}

\cortext[cor1]{Corresponding authors}

\affiliation[a]{Colorado School of Mines, Department of Physics, Golden, CO, USA}
\affiliation[b]{Laboratoire Univers et Particules de Montpellier, Montpellier, France}
\affiliation[c]{University of Chicago, KICP, Chicago, IL, USA}
\affiliation[d]{University of Iowa, Department of Physics and Astronomy, Iowa City, IA,  USA}
\affiliation[e]{NASA Goddard Space Flight Center, Greenbelt, MD, USA}
\affiliation[f]{Columbia University, NY 10027, USA}

\begin{abstract}
Space missions offer unique opportunities for studying ultra-high-energy (UHE) cosmic rays and neutrinos by leveraging secondary emissions generated by extensive air showers (EAS) resulting from their interactions with the Earth's atmosphere or crust. Detecting UHE neutrinos associated with transient sources holds great potential for unraveling the origins of UHE cosmic rays and the physical processes driving their production. Stratospheric balloon missions, illustrated recently by the Extreme Universe Space Observatory on a Super Pressure Balloon II Mission, serve as crucial precursors to space missions. Due to slewing abilities of their telescopes, they can perform follow-up observations of transient sources foreseen as potential candidates for detectable UHE neutrino emissions. Strategies tailored to stratospheric and space missions are essential for optimizing follow-up observations aimed at detecting these elusive neutrinos for the first time. To address this challenge, we have developed  flexible software dedicated to scheduling transient source observations. The software comprises three main modules: a listener module that aggregates alerts from existing alert systems to construct a comprehensive source database, an observability module that factors in the detection system's properties and trajectory to determine a list of observable sources during a specific time frame, and a scheduler module that prioritizes observations and proposes an optimized observation schedule. The initial release of the neutrino target scheduler software is tailored to the requirements of stratospheric balloon missions, with mock observation examples provided for various flight scenarios. This version will be employed for the upcoming POEMMA Balloon with Radio flight in 2028. Future developments will extend its capabilities, and ensure its relevance for various types of missions.
\end{abstract}


\end{frontmatter}

\section{Introduction}

The past decades have seen the development of a flourishing landscape for multi-messenger and transient astronomy. Important milestones have been achieved with the observation of astrophysical high-energy (HE) neutrinos \citep{IceCube:2013cdw}. In particular, there are hints for coincidence of HE neutrinos with several categories of transient sources, such as blazar flares \citep{IceCube:2018dnn, IceCube:2018cha, 2019ATel12967....1T, 2020A&A...640L...4G}, and with established or probable tidal disruption events \citep{Stein:2020xhk, Reusch:2021ztx, vanVelzen:2021zsm}. In addition, there are detections of steady HE neutrino fluxes from the active galactic nucleus NGC~1068 \citep{IceCube:2022der} and from the Galactic plane \citep{IceCube:2023ame}. Recently, the first ultra-high-energy (UHE) cosmic neutrino has been observed by KM3NeT \citep{KM3NeT:2025npi}, in which a muon with energy $120^{+110}_{-60}$ PeV was detected. With the flux needed to produce one such event in KM3NeT apparently inconsistent with existing IceCube and Auger upper bounds on a diffuse neutrino flux, explanations have turned to transient UHE neutrino sources (see, e.g., refs. \citealp{Neronov:2025jfj, Fang:2025nzg}) or more exotic solutions using physics beyond the standard model (see, e.g., refs. \citealp{Anchordoqui:2025xug, Dev:2025czz, Farzan:2025ydi, Brdar:2025azm, Murase:2025uwv}).

HE to UHE neutrinos, as secondary products of the interaction of HE to UHE cosmic rays (CRs), are essential probes for the identification of the sources of UHE cosmic rays \citep{Guepin:2022qpl}. In addition, the observation of gravitational waves and multi-wavelength photons from a binary neutron star merger GW170817 \citep{LIGOScientific:2017zic} opens exciting prospects to study the physics of compact objects through multi-messenger observations. These perspectives are further enhanced by the advent of powerful instruments, on the one hand with the detection of an abundance of transient sources, for instance with the Vera Rubin Observatory (VRO) \citep{LSST:2008ijt}, and on the other hand with instruments designed for multi-wavelength observations, for example the Space-based multi-band astronomical Variable Objects Monitor (SVOM) \citep{Wei:2016eox}. Finally, with gamma-ray instruments aiming towards higher sensitivities and energies, e.g., the Cherenkov Telescope Array (CTA) \citep{Maier:2019afm}, the High-Altitude Water Cherenkov Observatory (HAWC) \citep{Goodman:2007zz} or the Large High Altitude Air Shower Observatory (LHAASO) \citep{LHAASO:2019qtb}.

In this landscape, the JEM-EUSO program (Joint Experiment Missions for Extreme Universe Space Observatory) \citep{Abe:2023lne,Plebaniak:2025ayq} aims at developing missions to detect UHE particles (cosmic rays, and potentially neutrinos and photons) operating in the high atmosphere and in space. Its primary scientific objectives encompass unraveling the origin and composition of UHE cosmic rays, and discovering very-high-energy (VHE) ($>20$~PeV) neutrinos originating from astrophysical sources. The previous missions EUSO-balloon \citep{Adams:2022oko} and EUSO-SPB1 \citep{JEM-EUSO:2023ypf} paved the way towards high altitude UHE cosmic-ray observation. The Extreme Universe Space Observatory on a Super Pressure Balloon II Mission (EUSO-SPB2) was launched from Wanaka, New Zealand on May 13th 2023 \citep{Eser:2023lck,Adams:2025owi}. Quickly after its launch, EUSO-SPB2 reached its nominal altitude of $33$~km. Despite a short flight (2 days) due to a leak in the balloon, EUSO-SPB2 demonstrated the capabilities of the instruments to operate in a sub-orbital environment. The POEMMA-Balloon with Radio (PBR), with a 2028 launch date, will continue this observation program \citep{Battisti:2024jjy,Eser:2025fcw,Adams:2026cpe}. 

EUSO-SPB2 was equipped with a fluorescence  Telescope (FT) \citep{Adams:2024gsj}, a Cherenkov Telescope (CT) \citep{Gazda:2023tbw, Matamala:2023iru} and an Infrared (IR) camera \citep{Diesing:2023wcq}. The FT aimed at detecting UHE CRs, whereas the CT aimed at detecting CRs from above the limb \citep{Cummings:2023ypo, Fuehne:2023kap} and potentially VHE neutrinos from below the limb, through the fast beam of Cherenkov light emitted by extensive air showers (EAS). In the case of a VHE tau neutrino, the primary neutrino can interact with the Earth's crust and produce a secondary tau lepton. This particle can leave the Earth and decay in the atmosphere and thereby initiate an EAS.  To detect Cherenkov light emitted by EAS, the CT camera was equipped with 512 silicon photomultipliers (SiPM), covering a FOV of $6.4\degree \times 12.8\degree$ (vertical and horizontal).

An important capability of the CT was the ability to repoint, in zenith, to search for VHE neutrinos below the limb and VHE CR above the limb. The CT also had the ability to repoint in azimuth, allowing for follow-up observations of alerts associated with transient sources \citep{Heibges:2023yhn,Heibges:2025yvw}. To fully exploit these capabilities, we developed a Target-of-Opportunity (ToO) observation program prior to the SPB2 flight. This, in turn, required a dedicated open-source software tool to support the specific demands of this program.

A number of Python packages for scheduling astronomical observations are available, such as astroplan \citep{2018AJ....155..128M}, SCOPES (\url{https://github.com/nicochunger/SCOPES}), and the Observatory Control System (\url{https://github.com/observatorycontrolsystem}). However, these tools are primarily designed for ground-based observatories and thus offer limited support for moving detectors, extended FOV, or observations below Earth’s limb. While they include valuable features for specific contexts, such as SCOPES’ multi-telescope optimization capabilities or the Observatory Control System’s tools for submitting observation proposals, they do not address the unique challenges of balloon-borne observatories. For missions like EUSO-SPB2, scheduling observations is particularly complex due to uncertain and highly variable flight trajectories and observing conditions driven by unpredictable wind patterns. Although some automated solutions for airborne observational planning have been developed \citep[e.g.][]{2006smci.conf...80F}, they are either not publicly available or not adaptable to our specific constraints.

To address these limitations, we developed the Neutrino Target Scheduler (NuTS) software, specifically designed to schedule follow-up observations of transient sources. The NuTS software was developed as part of a general effort to expand $\nu$SpaceSim, an end-to-end simulation chain dedicated to the observation from high atmosphere or space of VHE-UHE particles producing EAS \citep{Krizmanic:2023pwf,Garg:2022ugd, NuSpaceSim:2023ims,Reno:2025tpw}. NuTS is built as a modular tool, allowing it to be easily adapted to a wide range of detector configurations beyond the specific case of the SPB2 or the PBR missions. NuTS is of particular interest for detectors with field of views covering only a fraction of the sky but able to repoint, as it can help optimizing their observation strategies.

In section~\ref{sec:structure}, we describe the general structure of the NuTS Software. In section~\ref{sec:schedules}, we describe the various strategies developed to schedule observations through examples. Conclusions and perspectives are described in section~\ref{sec:prospects}. In the Appendix, we provide additional information about: the online documentation in \ref{appendix:documentation}, the installation in \ref{app:intallation}, the usage in \ref{app:usage}, the data format in \ref{app:data-format}, field tests of the scheduler using starlight in \ref{app:test}, and information about observation schedules in \ref{app:app_schedules}.

\section{Structure of the software}\label{sec:structure}

\subsection{Overview}

\begin{figure*}[t]
    \centering
    \includegraphics[width=.98\textwidth]{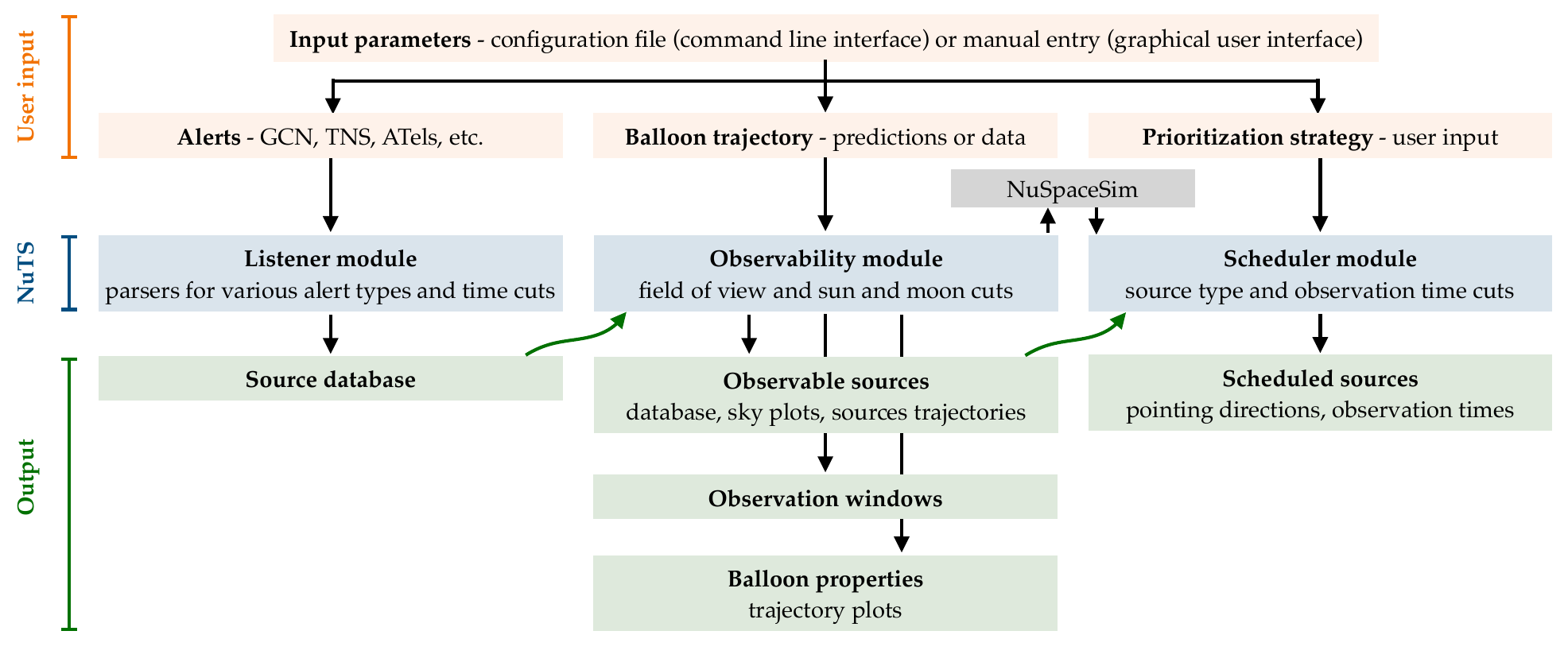}
    \caption{Main components of NuTS. The inputs to the listener module are alerts from GCN, TNS, Atels, etc. Collected alerts and a steady source catalog determine the source database. The balloon trajectory predictions or data determine the observation time window via the observability module. The scheduler module combines the source database and detector FOV to schedule sources, providing telescope pointing directions and observation times. }
    \label{fig:NuTSComponents}
\end{figure*}

The NuTS software is a Python package comprising several independent modules incorporated in a GitLab distribution: a listener module, an observability module, and a scheduler module. The listener module, described in section \ref{subsec:listeners}, interfaces with alert systems to collects alerts and to build a comprehensive database of energetic transient sources \citep{Wistrand:2023mpb}. The observability module, described in section~\ref{subsec:observability}, convolves this database with the properties of the detection system (observation period, trajectory, field of view, and other observability requirements) to produce a list of observable sources. The scheduler module, described in section~\ref{subsec:scheduler}, prioritizes observations, using constraints from the detection method and informed choices from models, to determine a specific schedule for a given observation period \citep{Posligua:2023cdm}. These main components are illustrated in figure~\ref{fig:NuTSComponents}.

As the three modules often need to be run independently and on different timescales, we chose this specific modular design for NuTS. The listener often needs to be run independently and continuously, in order to constantly update the source catalog with new alerts. The observability module is often run only once before each observing window, and it can be used independently from the listener module if a source database in the correct format is provided as an input. Finally, the scheduler can be run several times to compare the output of several strategies to find the optimal one; similarly to the observability module, it can be run independently after providing a list of observable sources and their properties, in the correct format as an input. In the following subsections, we describe these modules in more detail.

We developed a command line interface and a graphical user interface to make the software accessible to a variety of users and in different contexts, for instance for bursts advocates\footnote{Burst advocates coordinate alert-related activities and decide on source follow-up strategies.} during operations of the instrument. After installation (see Appendix \ref{app:intallation}), the user starts by initializing the software with the command
\begin{lstlisting}[language=bash]
nuts init
\end{lstlisting}
and by creating a configuration file with 
\begin{lstlisting}[language=bash]
nuts make-config config.toml
\end{lstlisting}
Then the user can run the different modules. The listener module runs with 
\begin{lstlisting}[language=bash]
nuts listen config.toml -l <listener_option> 
\end{lstlisting}
where \verb|<listener_option>| is set to \verb|TNS| or \verb|GCN|. The observability and scheduler modules run together or separately with
\begin{lstlisting}[language=bash]
nuts run config.toml -o <option>
\end{lstlisting}
for which many options are available and described in \ref{app:usage}. When running NuTS, a directory is created using the path and name specified in the configuration file. This directory contains a log file with all information about the run, together with results and figures. Results are stored in cvs and json files and consist of lists of all sources, observable sources and scheduled sources, with the format described in \ref{app:data-format}.

\subsection{Listener module}
\label{subsec:listeners}

The listener module is designed to receive and filter alerts from various alert networks. To use the listener module, the user must register with the alert systems. \ref{appendix:alert} has additional details. Then the module can be operated from the command line or the graphical user interface, as explained in \ref{app:usage}. 

NuTS is designed as a real-time analysis tool that can assist in planning observation strategies for transient sources. The user can establish various criteria that determine when alerts are added to and removed from the source database, based on the alert type, on the known or most probable source type when available, and the time of detection. The alerts that pass our selection criteria are added to the source database in real-time. The listener module receives alerts from the Gamma-ray Coordination Network (GCN\footnote{Gamma-ray Coordinates Network, \url{https://gcn.gsfc.nasa.gov/}}), the Transient Name Server (TNS\footnote{Transient Name Server, \url{https://www.wis-tns.org/}}), and the Astronomer’s Telegram (ATels\footnote{Astronomer's Telegram, \url{https://astronomerstelegram.org/}}). GCN and TNS alerts are available in machine-readable formats. GCN alerts are received and filtered in real-time, using the GCN-Kafka client. For TNS alerts, there is generally a delay between the detection and the creation of the alert (from hours to days, sometimes weeks), due to the time required for the classification of the object (for instance with additional spectroscopic information). Periodic requests to the TNS database can be sent at a user-defined cadence, with an upper limit of 1 request per 2 minutes. ATels are not machine-readable and thus are hand-processed by the user. 

The alerts selected correspond to potential sources of VHE neutrinos, such as binary neutron star mergers (BNSs), blazar flares, tidal disruption events (TDEs), gamma-ray bursts (GRBs) and several types of supernovae (SNe).

Some alerts correspond to flares from known sources (for example, blazar flares) and some provide the most probable source type (for example, gravitational wave events). TNS alerts are updated when the transient source has been classified (mostly TDEs and various types of SNe). When the most probable source type is not provided in the alert, we consider the history of similar alerts and associate the alert with the most probable source type. For instance, we systematically classify Fermi-GBM and Swift-BAT alerts as GRBs. Each alert is thus associated with a tag corresponding to the most probable source type, which is later used to determine scheduling priorities.

Finally, outdated alerts are systematically removed from the database, setting maximum retention times based on the source or type of alert. This is done by continuously scanning the database and comparing the time of first observation of an alert with the user defined retention time for this alert type. NuTS keeps a record of all  alerts, facilitating backwards checking of alerts, but in the real-time mode, only the selected subset is processed for follow-up observations.

In addition, a default database is provided with the software, where steady sources that are promising candidate sources for the emission of VHE neutrinos are selected from several catalogs. We include several blazars from the Whole Earth Blazar Telescope (WEBT\footnote{The Whole Earth Blazar Telescope, \url{https://www.oato.inaf.it/blazars/webt/}}), 
and several starbust and Seyfert galaxies, for example, M 82 and Cen A. From TeVCat\footnote{TeVCat catalog for TeV Astronomy, \url{http://tevcat.uchicago.edu/}} \citep{2008ICRC....3.1341W}, we include the Galactic center, various types of BL Lacertae objects, namely high-frequency peaked BL Lacs (HBLs), intermediate BL Lacs (IBLs) and low-frequency peaked BL Lacs (LBLs), flat-spectrum radio quasars, Fanaroff-Riley class I objects, the massive stellar clusters Westerlund I and II, and the supernova remnant SNR G106.3+02.7. In addition, we add confirmed or suspected sources of HE neutrinos identified by IceCube, for example, TXS 0506+056 and NGC 1068 \citep{IceCube:2018cha, IceCube:2022der}. With these different components, a source database is produced by the software in a generic format (csv). Additional sources can be added by the user to the steady source database.

\subsection{Observability module}\label{subsec:observability}

The observability module determines the list of sources that can be observed within a given observation window, using the source database as input. For each observable source, the module computes the range of times and azimuths for which this source is visible. These quantities are then used by the scheduler module to construct the observation plan. The observability module is executed with the following command
\begin{lstlisting}[language=bash]
nuts run config.toml -o <option>
\end{lstlisting}
where \verb|<option>| is set to \verb|observability| or \verb|observations|. The first option uses a pre-existing database whereas the second recomputes a new database by merging databases obtained from the listener and/or user-provided entries, while removing outdated alerts. These execution sequences correspond to functions defined in the file \verb|compute.py| and can be easily modified in developer mode. A validation test of the observability module was performed by measuring the altitude of various stars and comparing these measurements with NuTS predictions. The results of this test are described in \ref{app:test}. In the following, we describe the algorithm responsible for determining observable sources, and we detail the required elements entering this algorithm.

\subsubsection{Observability algorithm}

The main steps required to determine source observability are described below.

\begin{itemize}
    \item Database loading: Load the catalog of candidate events with sky coordinates (e.g., neutrino alerts, gamma-ray bursts, gravitational-wave localizations). 

    \item Trajectory \& detector geometry: Retrieve the planned balloon trajectory (altitude, latitude/longitude, time, see section~\ref{sec:traj}) and the detector’s FOV (azimuth and altitude extents, see section~\ref{sec:fov}). The input parameters for the FOV and the detector orientation are optimized for neutrino searches below the Earth's limb. From those, compute the instantaneous pointing geometry as a function of time.

    \item Illumination constraints: Apply Sun and Moon exclusion criteria, that is remove time segments where the Sun or Moon illumination prevents valid Cherenkov-signal detection (e.g., too bright sky background or angular proximity, see section~\ref{sec:sunmoon}).

    \item Visibility computation: For each candidate source, compute the time intervals during which the source falls within the FOV given the balloon position and pointing geometry, taking into account the rotation of the sky, horizon limits, and payload motion. For point sources, their coordinates determine directly their observability, while for extended sources or poorly localized sources, a specific algorithm (see section~\ref{sec:poorlyloc}) is used to identify the highest-probability regions for observation.

    \item Selection of observable windows: Identify contiguous time blocks where the source remains below the horizon and within the FOV, and free of Sun/Moon exclusion criteria. Save for each source the earliest start time, the latest end time, and the total available observation duration.

    \item Output generation: Write out into output files source tables for all processed sources and for observable sources. When applicable, save their start/end times and the pointing parameters. Several output formats are produced, and a specific format has been created for sensitivity computation through the interfacing with $\nu$SpaceSim (see \ref{sec:output_tsprpst}).

\end{itemize}

By combining flight geometry, detector field-of-view and illumination constraints in a unified framework, this module ensures that only observable sources are forwarded for scheduling.

\subsubsection{Balloon trajectory}\label{sec:traj}

\begin{figure}[ht]
    \centering
    \includegraphics[width=.49\textwidth]{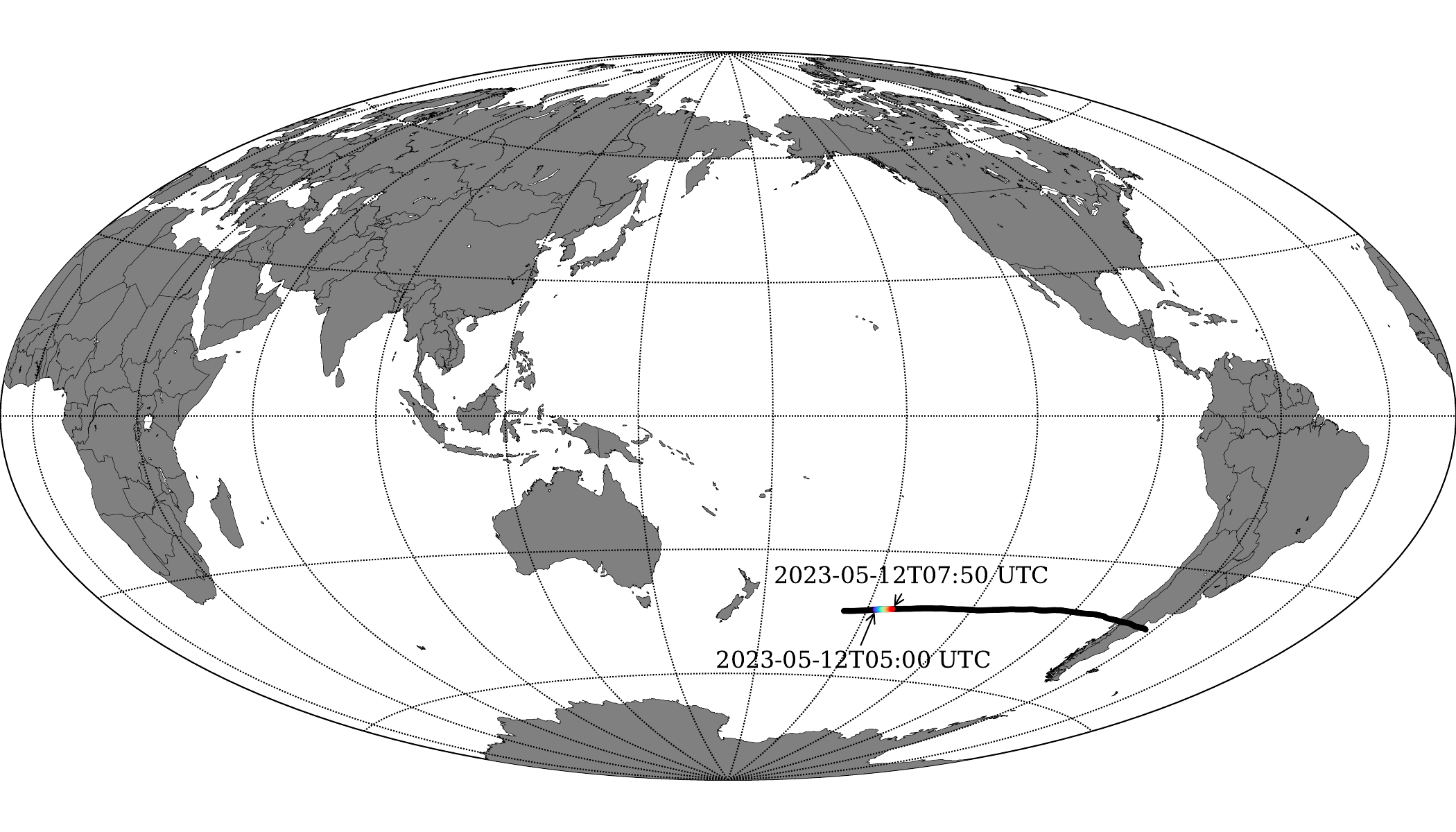}
    \includegraphics[width=.48\textwidth]{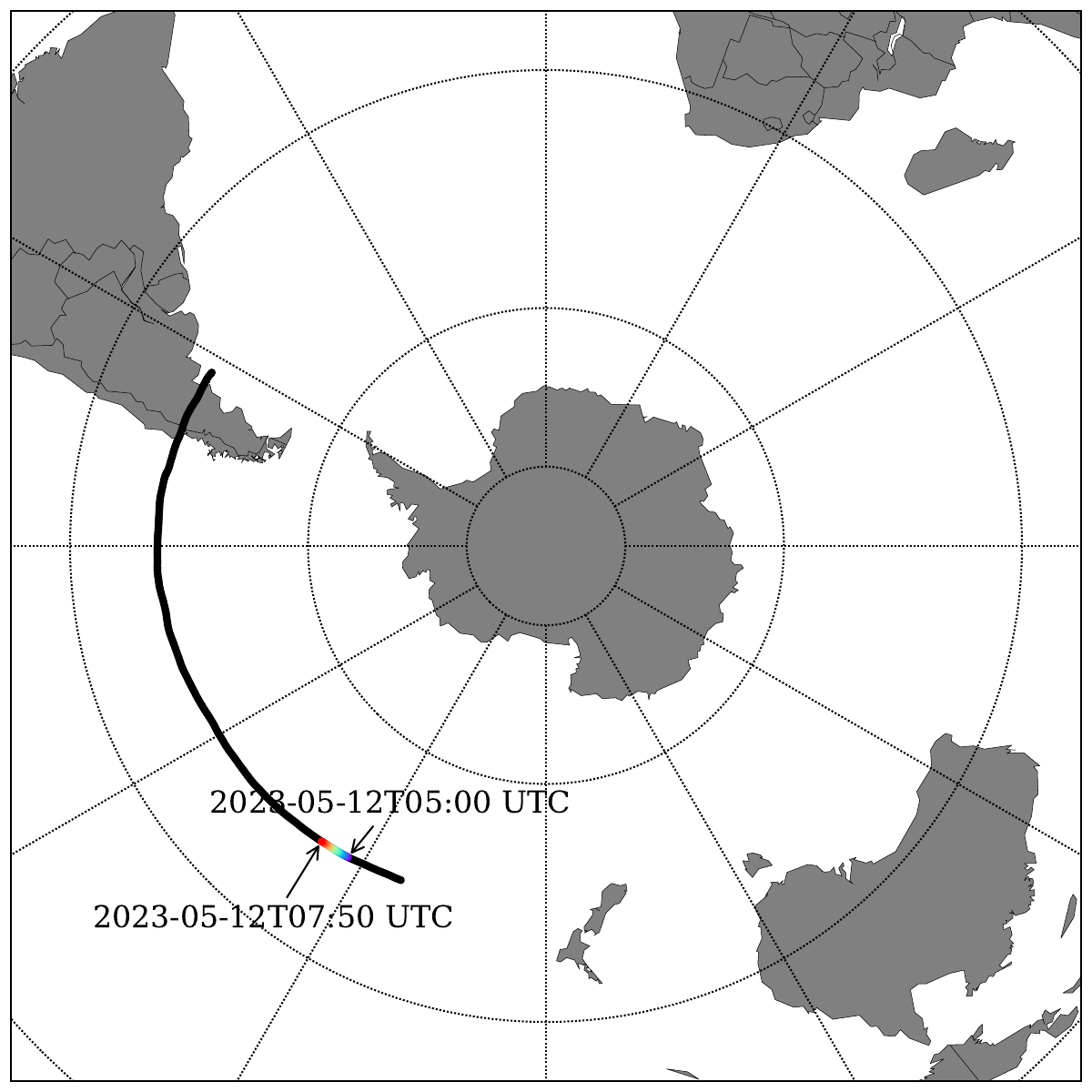}
    \caption{Trajectory of the SuperBit balloon. We show the trajectory starting from May 12, 2023 during $48$~hours (black) and during the observability window on May 12, 2023 (from purple to red). The full SuperBit mission lasted $\sim 45$ days \citep{Gill:2024mqb}.}
    \label{fig:DetectorTrajectory}
\end{figure}

The trajectory of a payload impacts the observable fraction of the sky, as it influences the evolution of the instantaneous field of view, and the duration of the observation window. In the context of the EUSO-SPB2 mission, we initially focused on accounting for the trajectory of a 18MCF super-pressure balloon taking off from Wanaka (New Zealand). These super-pressure balloons are expected to reach high altitude winds circling around Antarctica. Their nominal float altitude is 33~km, with daily variations in float altitude much less than 1~km. A super-pressure balloon flight may have up to a 100-day duration.

\begin{figure*}[ht]

    \centering
    \includegraphics[width=.48\textwidth]{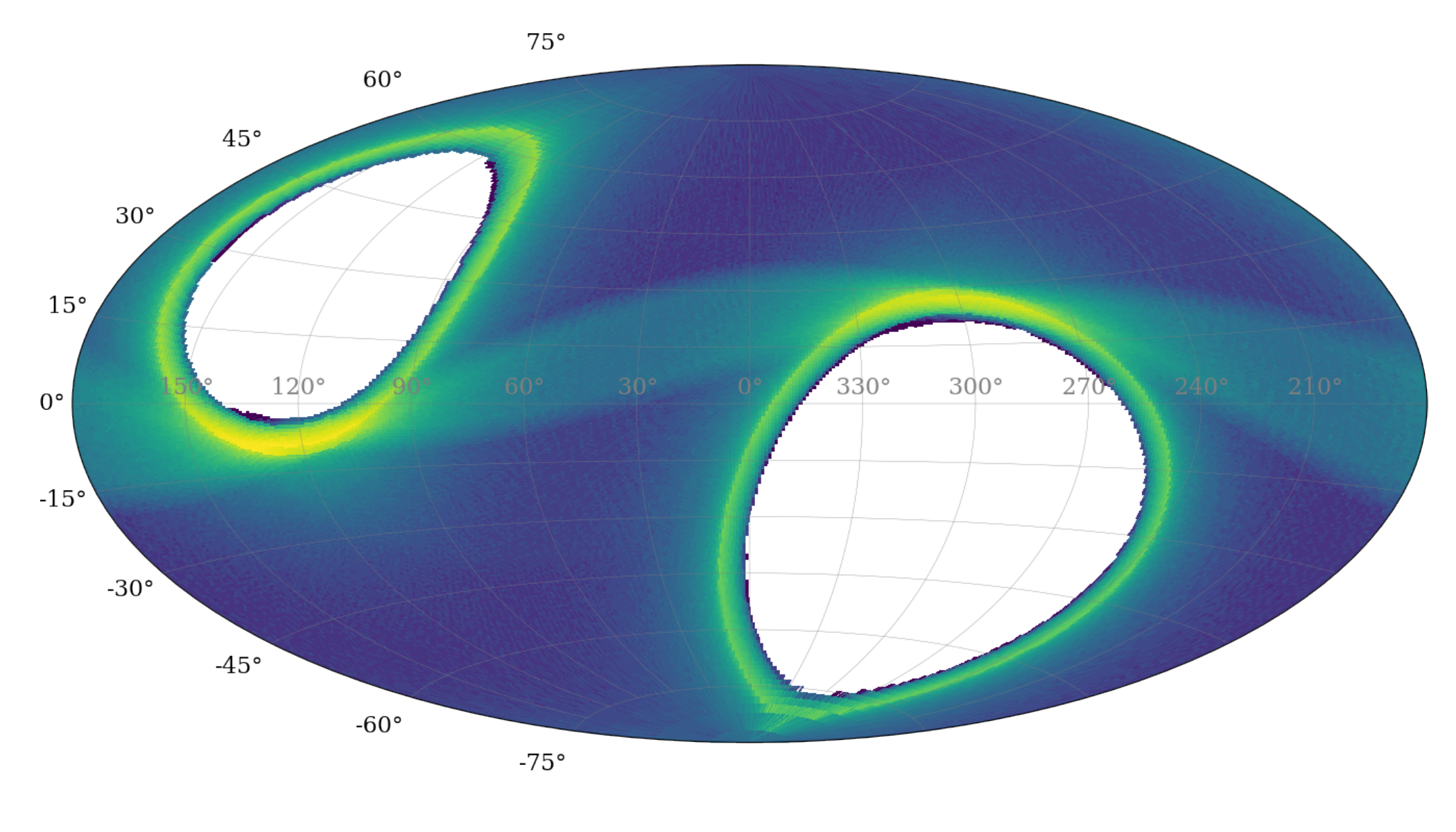}
    \includegraphics[width=.48\textwidth]{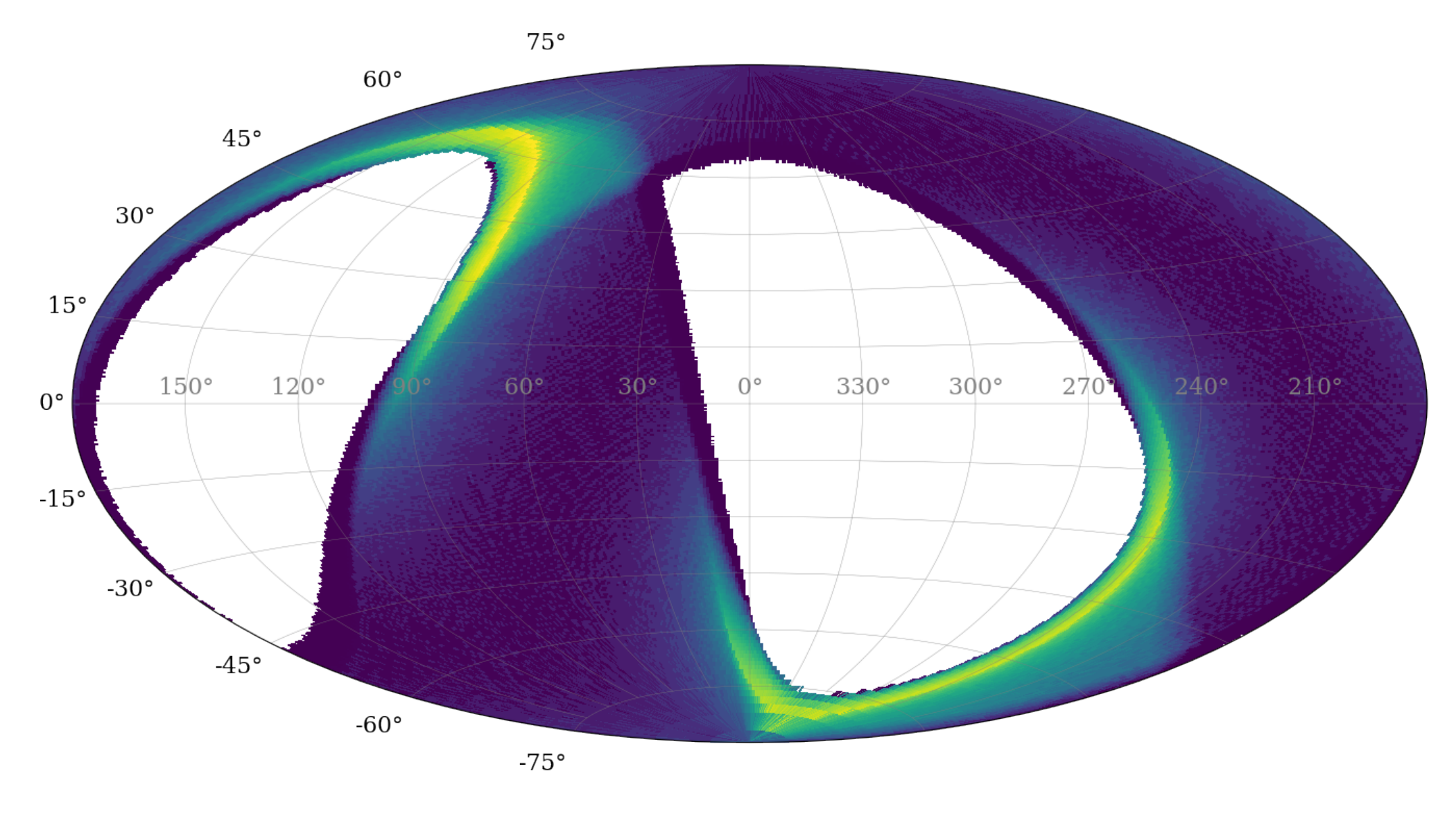}
    \includegraphics[width=.8\textwidth]{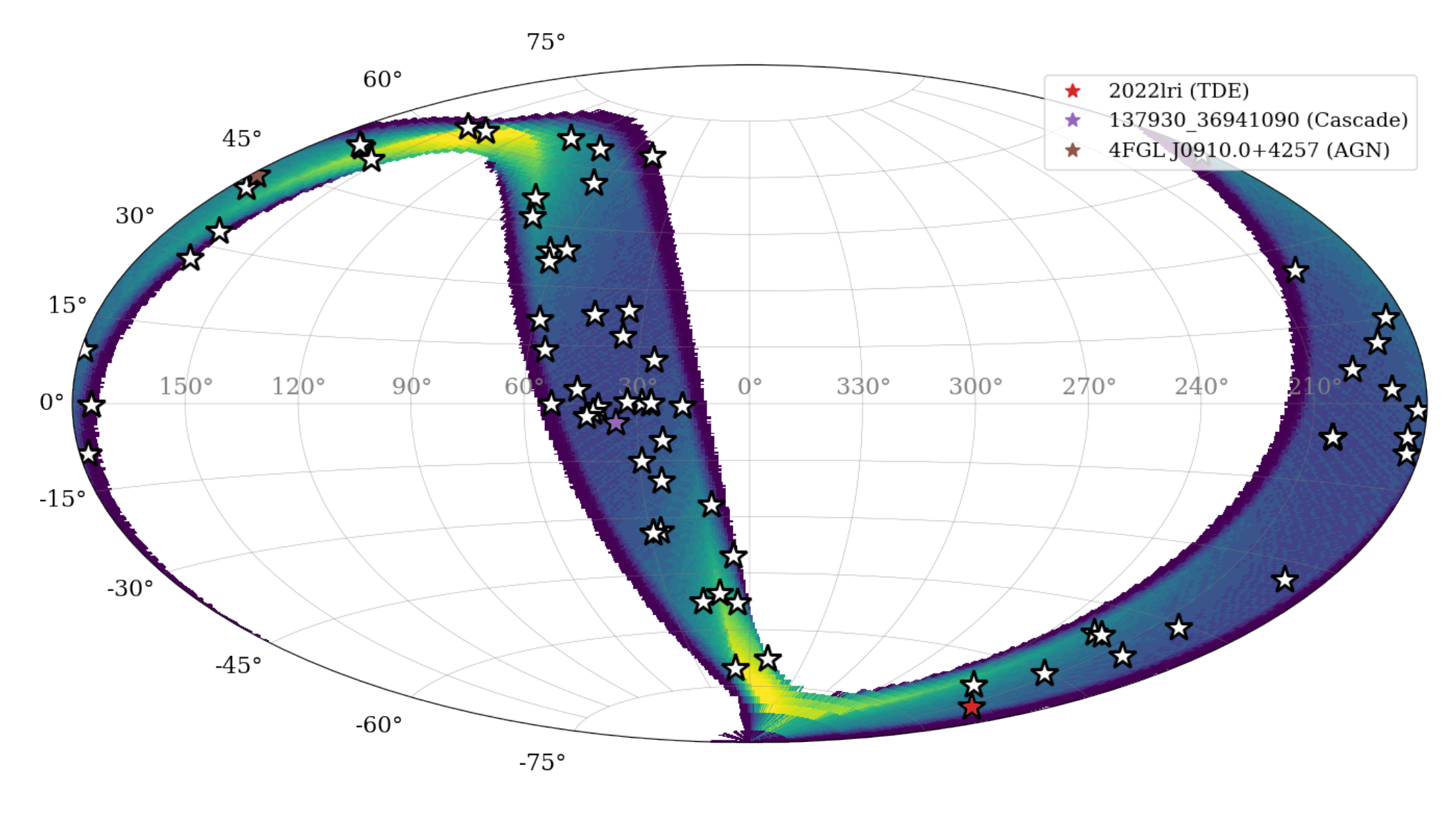}
    \caption{Geometrical acceptance (see text for definition) in Galactic coordinates, for similar conditions as figure~\ref{fig:DetectorTrajectory}, and for a time window of $24~$hours centered on the observability window of May 12, 2023. We account for the observability constraints from the field of view (top left panel), we add the constraint from the Sun (top right panel) and the Moon (bottom panel). For each panel, we normalize the acceptance by its maximum value, and we set a logarithmic colorscale in the range $0.1 - 1$. On the bottom panel, we also show the observable sources (black and white stars) from a test catalog, and the scheduled sources (black and colored stars).
}\label{fig:SkyMap}

\end{figure*}

The trajectory input files can come in multiple formats, so we developed several routines dedicated to extracting the trajectory data. For instance, during the EUSO-SPB2 mission, the Columbia Scientific Balloon Facility (CSBF) provided us every day with KML files\footnote{\url{https://developers.google.com/kml/documentation/}} containing the balloon's projected latitude, longitude, and altitude for the next three days in $6\,{\rm h}$ bins. We also give the option of defining a constant position, with fixed latitude, longitude and altitude. Two other formats are available options, a log file format that describes a trajectory after the end of the mission, and a csv file format that describes a simulated trajectory. The user-developer can define a custom parser for a specific format using the examples given in the software.

After being processed by the parser, the trajectory vectors are interpolated linearly to determine the balloon positions at times required by the software. Several figures are produced while running the software to visualize and cross-check the trajectory of the detector, as shown in figure~\ref{fig:DetectorTrajectory} for a snapshot of the SuperBit \citep{Gill:2024mqb} trajectory. The figure shows the SuperBit balloon trajectory during 48 hours. The observation window on May 12, 2023 was short, between 5:00UTC-7:50UTC. The portion of the trajectory on this day during the observation window is shown with colors. 

The software is designed to be applicable for any type of detector and trajectory. It will eventually be adapted to satellites and ground-based telescopes.

\subsubsection{Field of view}\label{sec:fov}

Currently, a rectangular detector field of view is implemented in NuTS. More complex structures are left for future developments. The field of view of the detector is defined in the configuration file by two parameters, its extension in altitude $\Delta\alpha$ and azimuth $\Delta\phi$. As an example, in the case of the EUSO-SPB2 mission, the CT extends $\Delta\alpha = 6.4^\circ$ in altitude and $\Delta\phi = 12.8^\circ$ in azimuth. We use these values of $\Delta\alpha$ and $\Delta\phi$ in the examples below.

The parameter $\Delta\alpha_{\rm off}$ describes the offset of the detector orientation in altitude, defined with respect to the detector's local horizontal. The offset can be set to the Earth's limb to search for neutrinos. In this configuration, the detector is oriented so that the upper part of the field of view corresponds to the Earth's limb. In this case, as an example, $\Delta\alpha_{\rm off} \simeq 5.8\degree$ is obtained for a balloon at $33~$km.

The properties of the field of view are used to identify which sources can be observed during the observation window, and in general, the observable fraction of the sky. This corresponds to a purely geometrical criterion, with a flat acceptance between the upper and the lower end of the field of view.

\subsubsection{Sun and Moon illumination}\label{sec:sunmoon}

The observation of optical Cherenkov radiation requires a dark sky. As the software is initially designed to schedule observations with the CT on board of the EUSO-SPB2 and PBR missions, we define observability criteria depending on the illumination of the Sun and the Moon.

Observations are possible when the Sun and Moon have set, or when the Moon's illumination is below a given threshold. The default altitude thresholds for the Sun and the Moon are added to $\Delta\alpha_{\rm off}$, thus by default defined with respect to the Earth's limb, with $\Delta\alpha_{\rm sun} = -18^\circ$  (astronomical night) and $\Delta\alpha_{\rm moon} = -1^\circ$ (Moon set below the Earth's limb, accounting for refraction effects). The default threshold for Moon illumination is $0.05$. This cut is based on radiative transfer simulations of the transmission of sunlight in the atmosphere, scaled to estimate the scattered moonlight. Scaling factors include the distances between the Sun and the Moon and the Moon and the Earth, the Moon’s albedo, and the Earth’s albedo. The intensity of the scattered moonlight is required to be less than $10\%$ of the airglow background estimate. These default threshold values can be modified by the user in the configuration file.

Transient neutrino sources must be below the Earth's limb to be observable. In figure~\ref{fig:SkyMap}, we show the observable portion of the sky, for a $24$~hour time window centered around the observation window of May 12, 2023, for a mock flight using the trajectory of the SuperBit payload, with the EUSO-SPB2 FOV. This figure allows comparisons of the observable portion of the sky computed by considering only the constraint from the FOV (upper left panel), and by adding the Sun constraint (upper right panel), and finally adding the Moon constraints (lower panel). In this situation, the Moon illumination is high, and observations are only possible when the Moon is set. The range of colors corresponds to normalized observation times in a logarithmic scale, where yellow indicates the largest observation times. White regions are not observable on this day.

\subsubsection{Poorly Localized Sources}\label{sec:poorlyloc}

Some transient candidates, notably gravitational-wave (GW) events, are not localized to a single sky position but instead to extended probability regions covering tens to thousands of square degrees. In contrast to well-localized point sources, these events require a dedicated procedure to determine their observability and to compute an optimal pointing strategy for the telescope.

\begin{figure*}[t]
    \centering   \includegraphics[width=0.48\textwidth]{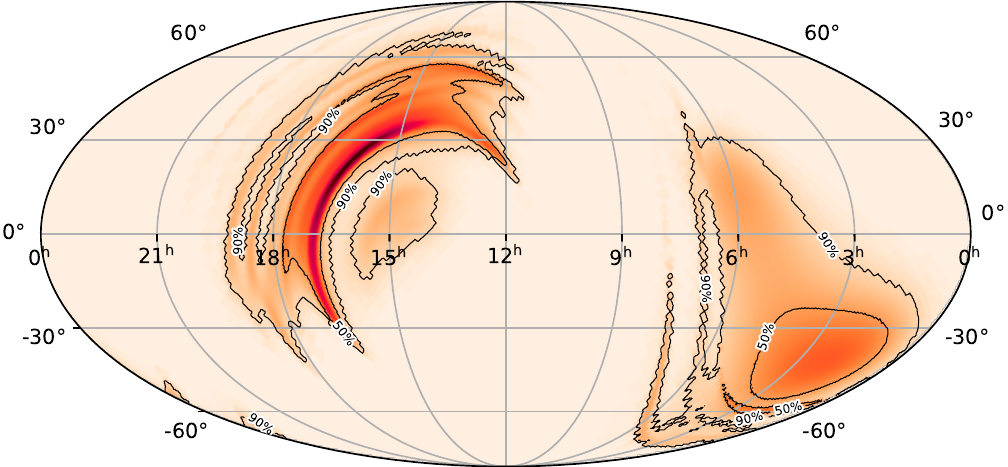}\vspace*{0.3cm}
    \includegraphics[width=0.48\textwidth]{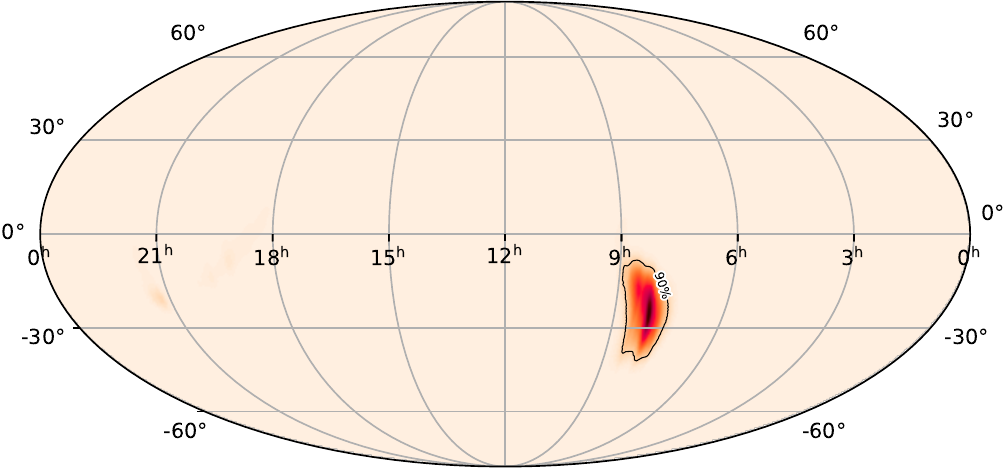}
    \caption{Two skymaps for GW events, respectively S190425z (left) and S240422ed (right) with their 90\% confidence regions indicated. These figures are generated by the ligo.skymap Python package (\url{https://lscsoft.docs.ligo.org/ligo.skymap/}).}\label{fig:GWskymap}\includegraphics[width=.49\textwidth]{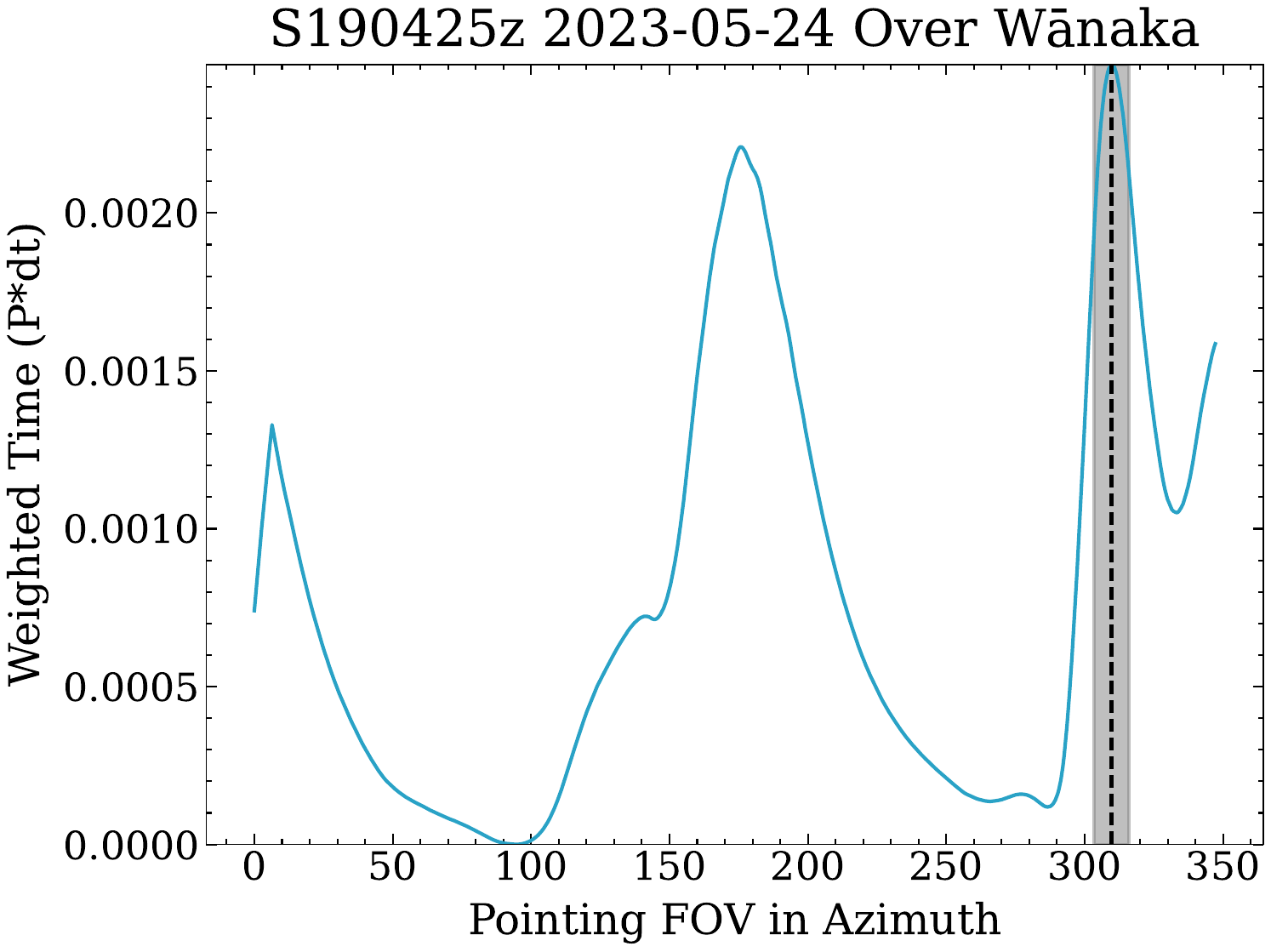}\includegraphics[width=.49\textwidth]{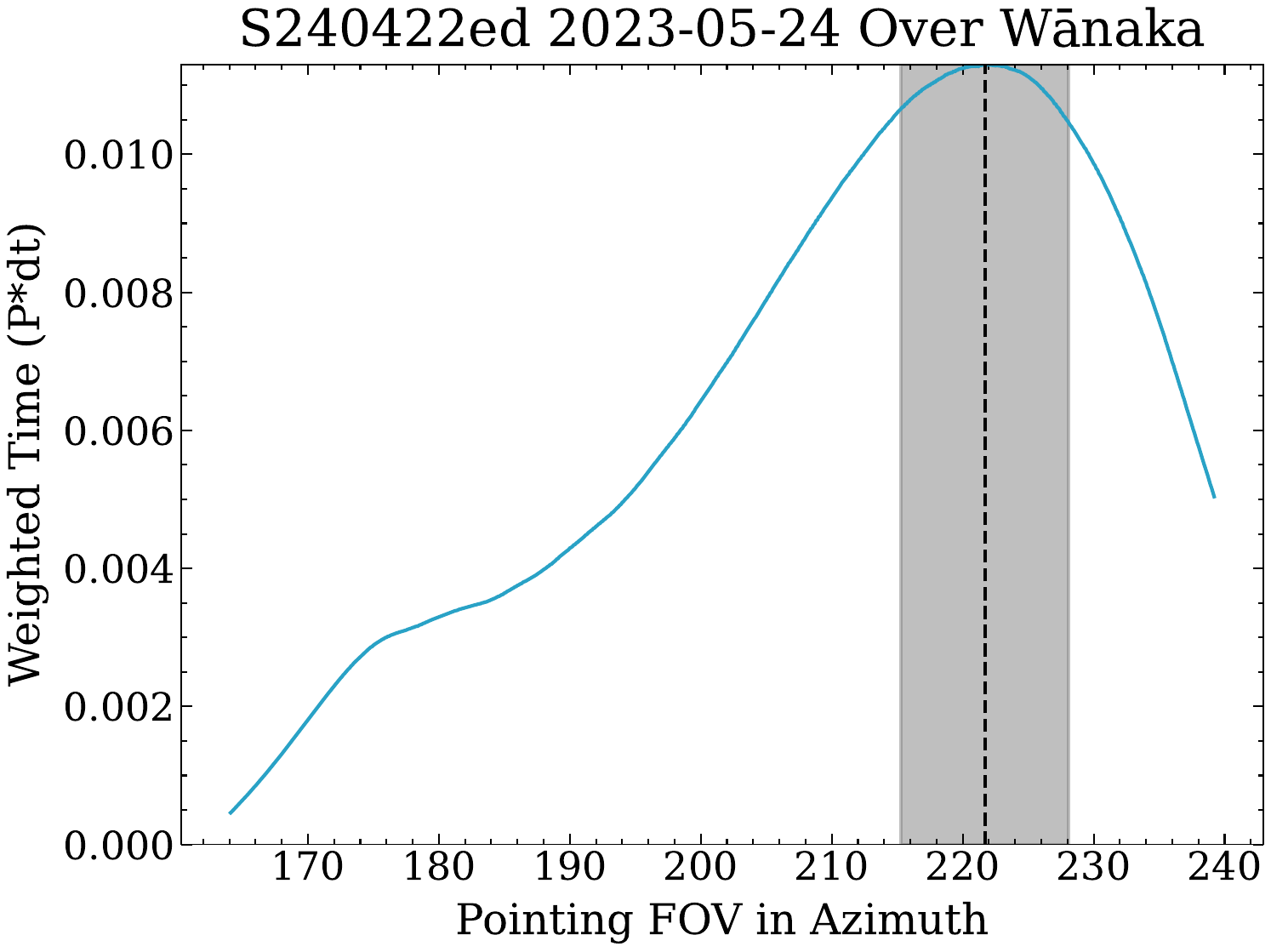}

    \caption{The weighted time (localization probability times pixel time in the FOV) for the GW events S190425z (left panel) and S240422ed (right panel), as a function of azimuthal pointing for EUSO-SPB2 on May 24, 2024 at a location over Wanaka, NZ. The azimuthal FOV is shown in grey. 
    }\label{fig:GWresult}
\end{figure*}

Figure~\ref{fig:GWskymap} illustrates two representative cases. S190425z (left), detected by only two GW interferometers (LIGO Livingston and Virgo), exhibits a broad 90\% confidence region (of order 10,000 deg$^2$), whereas S240422ed (right), observed by three detectors (LIGO Hanford, LIGO Livingston and Virgo), is localized more tightly, $\sim 400$ deg$^2$ for the 90\% confidence region. The first event is most likely a binary neutron star merger, whereas the second one, a black hole-neutron star merger.

Strategies for follow-ups of GW alerts are telescope dependent. Electromagnetic follow-ups with optical telescopes are effectively done by instruments that have fast slew times and short exposure times so that multiple telescope pointings can tile the localization regions. Large FOV telescopes such as the Rubin telescope (47 deg$^2$) are advantageous.  Even larger FOV high energy instruments and  radio telescopes can cover larger portions of GW localization regions with each exposure. Examples of projected rapid GW follow-up response times appear in, for example, \cite{Chu:2015jxa}. Whereas electomagnetic follow-ups of GW alerts can be done on the full observable sky, neutrino source follow-ups may be more constrained: the zenith pointing is fixed to be below the limb of the Earth and azimuthal slewing may not be rapid. In the case of EUSO-SPB2 and PBR, azimuthal slew times are of the order of $5-15$ minutes. The GW follow-up strategy for neutrino telescopes on stratospheric balloon missions is to find the azimuthal pointing to optimize the angular area overlap of the telescope's neutrino FOV with the localization region as well as to optimize observation times, a good proxy for acceptance, for the overlap region. In this approach, by choosing a single optimal azimuthal pointing direction, time is devoted to observing rather than slewing the telescope.

GW localization skymaps, distributed through the Gravitational-Wave Candidate Event Database (GraceDB\footnote{LIGO/Virgo/KAGRA Public Alerts, \url{https://gracedb.ligo.org/superevents/public/O4/}}), are provided as HEALPix maps. NuTS reads these FITS files using the \texttt{healpy} package and extracts each pixel's sky coordinates together with its associated localization probability. The angular area of each pixel depends on the map; for example, for S190425z, the pixel area is $5.25\times 10^{-2}$ deg$^2$, while for S242422ed, the pixel area is a factor of 0.25 smaller. We downgrade the resolution of the original skymap to approximately 1 deg$^2$ by default (\texttt{healpy} value of \texttt{nside=64}) once the sky map has been read in with \texttt{healpy}. The factor with which the software down-samples can be user defined in the config file. Down-sampling to $\sim 1$ deg$^2$ decreases the total number of pixels in the localization regions
significantly since the regions are so large, as noted above. The algorithm treats every HEALPix pixel inside the chosen confidence region (typically 90\%) as an independent ``point-source'' candidate. For S190425z, the 90\% confidence region is represented by $\sim 10^4$ points, while S240422ed is represented by $\sim 400$ points. For each of these points:
\begin{itemize}
    \item its right ascension and declination are passed to the observability module;
    \item the observable time intervals are computed using the balloon trajectory, detector field of view, and Sun/Moon constraints;
    \item the pixel trajectory across the detector's zenith-pointed FOV is evaluated at the user-defined time resolution (e.g.,\ 1\,min);
    \item for all observable intervals, the duration spent within the FOV extension in altitude $\Delta\alpha$ is recorded.
\end{itemize}

To determine an optimal pointing direction for the telescope, the algorithm performs a scan over azimuth in small increments (typically $0.1^\circ$) across $360^\circ$. For each trial azimuthal orientation, the following quantity is computed:
\begin{equation}
    t_{\mathrm{w}}
    = \sum_{i \in \mathrm{pixel}} 
      p_i \, \Delta t_i,
\end{equation}
where $p_i$ is the localization probability of the sky coordinate of pixel $i$, and $\Delta t_i$ is the time that the pixel $i$ sky coordinate spends inside the field of view. This ``weighted time'' serves as an estimator of the fraction of the probability region that can be observed and how long it remains observable. The azimuthal angle that maximizes $t_{\mathrm{w}}$ is selected as the optimal pointing for the event. The earliest start time and latest end time over all pixels entering the field of view determine the corresponding observation window.

Figure~\ref{fig:GWresult} shows examples of the resulting weighted-time profiles for S190425z and S240422ed as a function of azimuth pointing, given EUSO-SPB2 CT positioned at 33 km above Wanaka, NZ. The shaded band shows $\Delta \phi = 12.8^\circ$, the range of $\phi$ (azimuth extent) for EUSO-SPB2's CT. The peak structure reflects the geometry of the probability region: extended or multimodal sky maps produce multiple local maxima, illustrating that near-optimal pointings may exist besides the global maximum. While the scheduler currently uses only the optimal pointing and observation window, these visualizations provide diagnostic information on the structure of the localization region and can assist users in manually selecting alternative or additional scientifically relevant pointing strategies.

\subsubsection{Output for sensitivity computation}\label{sec:output_tsprpst}

NuTS supports intermediate output of the source locations as they cross through the observable range in the sky. It supports output into HDF5, FITS, and npy/npz output files. This functionality allows for both saving the source locations and passing them to other analysis tools, such as  $\nu$SpaceSim, as well as performing fast scheduling updates that rely on previously calculated locations of the source. 

The input-output is handled through a dedicated Target-Source-Position-Relative-Point-Source-Tensor (TSPRPST) class. This class contains information about the detector location as a function of time and on a source-by-source level, the time, altitude, and azimuth of the observable source while it is in the observable region. It also supports multiple observation windows in a single file.

\subsubsection{Adapting the observability module to other detectors}

Several elements in the observability module were developed specifically for the EUSO-SPB2 and PBR missions:
\begin{itemize}
    \item Balloon trajectory: for detectors on the ground a constant detector location can be chosen, while for satellites, additional developments will be required.
    \item Illumination constraints: these are determined by the UHE neutrino detection method used in the EUSO-SPB2 and PBR missions (the CT camera). For other detection methods such as radio detection, these constraints can be toggled off in the input parameters.
    \item Visibility computation: this part is determined by the extension of the detector's FOV above and below the horizon, which can be changed in the input parameters. At UHE energies, the probability of detecting neutrinos diminishes sharply as the observation angle moves further below the horizon. As a result, most detection methods prioritize Earth-skimming trajectories.
\end{itemize}

As explained above, most of these elements specific to the EUSO-SPB2 and PBR missions can be adjusted or toggled off in the input parameters. When in depth modifications are required, users are encouraged to contact the developers.

\subsection{Scheduler module}\label{subsec:scheduler}

The list of observable sources can be extensive, with up to tens or hundreds of observable sources per night. However, only a few of those sources can be scheduled for observation because the observation time is limited, and in some cases, only a few re-pointings can be performed during each observation run. For instance, for the EUSO-SPB2 mission, about five re-pointings in azimuth could be scheduled every night. To efficiently select the most promising sources for observation, a scheduling strategy is implemented and managed by the scheduler module. The scheduler can be executed using the following command
\begin{lstlisting}[language=bash]
nuts run config.toml -o <option>
\end{lstlisting}
where \verb|<option>| is set to \verb|schedule|, or other options \verb|obs_sched| and \verb|all| that allow the user to run at once the observability and scheduler modules. We first describe the algorithm responsible for scheduling individual sources, followed by the procedure that defines the sequence in which they are considered.

\subsubsection{Scheduling algorithm}

According to a defined prioritization strategy, the scheduler processes sources sequentially. The algorithm implemented in the source file \verb|schedule.py| determines, for each source, its scheduling feasibility as well as the optimal observation times and detector orientations. The core procedure executes the following steps:

\begin{itemize}
    \item Initialization: A local schedule is created for the candidate source, defining a grid of available times from the global configuration and initializing all slots as unallocated.

    \item Visibility determination: The method computes the intervals during which the source is observable within its specified observation window. Source coordinates are transformed into the detector reference frame to extract the visible time samples.

    \item Segmentation of observation periods: Contiguous visibility intervals are identified by detecting time discontinuities larger than a predefined threshold. Only sufficiently long, continuous windows are retained for scheduling.

    \item Geometrical and operational checks: For each potential window, azimuthal motion is compared against the detector’s field of view to ensure that the observation remains within operational limits. The algorithm then adjusts the observation start time to include telescope repointing, and verifies that the minimal observation duration and maximum number of scheduled sources are not exceeded.

    \item Scheduling decision: If all criteria are satisfied, the corresponding time slots are marked as allocated. The source is added to the global schedule, storing the observation period, pointing direction, and detector configuration.
\end{itemize}

The method iterates through all observation windows until a complete schedule is found or no feasible slot remains. 

As highlighted above, several parameters influence the observation schedule. The time buffer \verb|rotation_time| accounts for the time required to re-point between two observations. Its default value is $10$~min. The time threshold \verb|min_obs_time| accounts for the minimum time during which a source can be observed in order to be scheduled. Its default value is $15$~min.

\subsubsection{Prioritization strategies}

The prioritization strategies, implemented in the source file \verb|schedule_strategies.py|, define the order in which sources are considered for scheduling. They are controlled by parameters specified in the configuration file under \verb|settings.scheduler|:
\verb|strategy_priority_sort|, \verb|strategy_priority_max|, \verb|strategy_obstime|, and \verb|strategy_obsprev|.
These parameters correspond to three main prioritization criteria: source type, expected observation time, and previous observation history. The criteria can be activated independently or combined according to user preference.

\paragraph{Source type criterion} Not all astrophysical sources have equal potential to produce detectable VHE neutrino fluxes. When \verb|strategy_priority_sort| is enabled, sources are scheduled following a predefined ranking based on expected detectability, from highest to lowest priority:
\begin{enumerate}
    \item Galactic supernovae (SNe),
    \item (V)HE neutrino events (from IceCube and KM3NeT) and binary neutron stars (BNS) mergers,
    \item flaring blazars (BL Lacs and FSRQs),
    \item tidal disruption events (TDEs),
    \item gamma-ray bursts (GRBs),
    \item active galactic nuclei (AGN) other than blazars,
    \item extragalactic SNe,
    \item steady sources.
\end{enumerate}
This hierarchy follows multi-messenger studies on VHE neutrino source detectability \citep{Venters:2019xwi, Guepin:2022qpl}. The highest priority ranking is given to the rarest and most promising events for VHE neutrino detection (Galactic supernovae, (V)HE neutrinos, and BNS mergers). Transient source categories that have exhibited potential coincidences with HE neutrino events (blazar flares and TDEs) are next, followed by GRBs whose large electromagnetic output suggests high potential despite no confirmed neutrino associations. AGN and extragalactic SNe receive lower priority due to their uncertain VHE neutrino yields, and steady sources rank last since short balloon flights cannot achieve comparable sensitivity to current (V)HE neutrino detectors. Users can freely modify this ranking or exclude specific source types. The parameter \verb|strategy_priority_max| limits the number of top-priority categories for which all observable sources are scheduled. Beyond this threshold, only a representative subset of sources per class is considered, preventing oversampling of any single category and reflecting uncertainties in multi-messenger models.

\paragraph{Observation time criterion} For the EUSO-SPB2 and PBR missions, the observation times for each source can vary between 20\,mins and 1\,h\,20\,mins, typically. The minimum observation times are obtained for sources crossing the FOV vertically, and are thus limited by the vertical extension of the FOV. The maximum times are related to sources displaying Earth-skimming trajectories, and in those cases several re-pointings are required to observe their full trajectories, as their observation is limited by the horizontal extension of the FOV. Given that observation time is a good proxy for acceptance, enabling \verb|strategy_obstime| favors sources with the largest available observation time within the mission window.

\paragraph{Previous observation criterion} Some categories of transient sources could produce VHE neutrinos during extended periods of time, from days to months. To increase the total exposure for such sources, the parameter \verb|strategy_obsprev| allows prioritizing objects previously observed or scheduled. The path to the prior schedule must be specified in the configuration file.

\subsubsection{Outputs and illustration}

\begin{figure}[ht]
    \includegraphics[width=.48\textwidth]{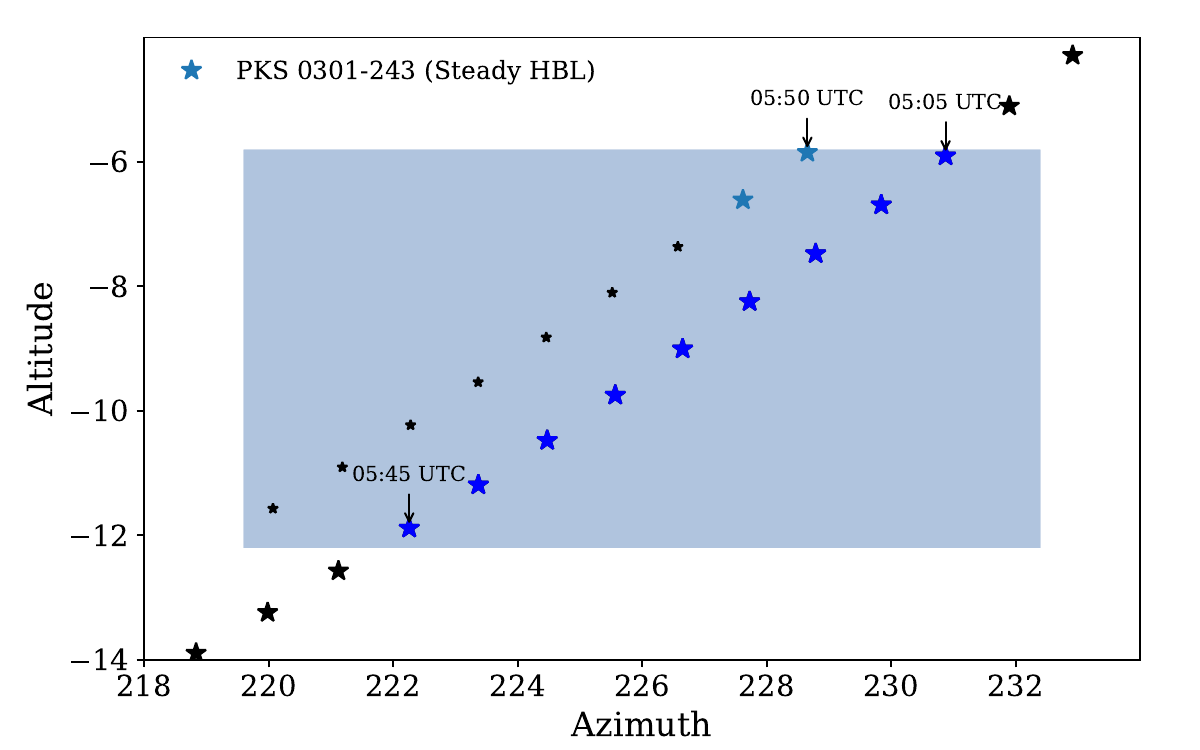}
    \includegraphics[width=.48\textwidth]{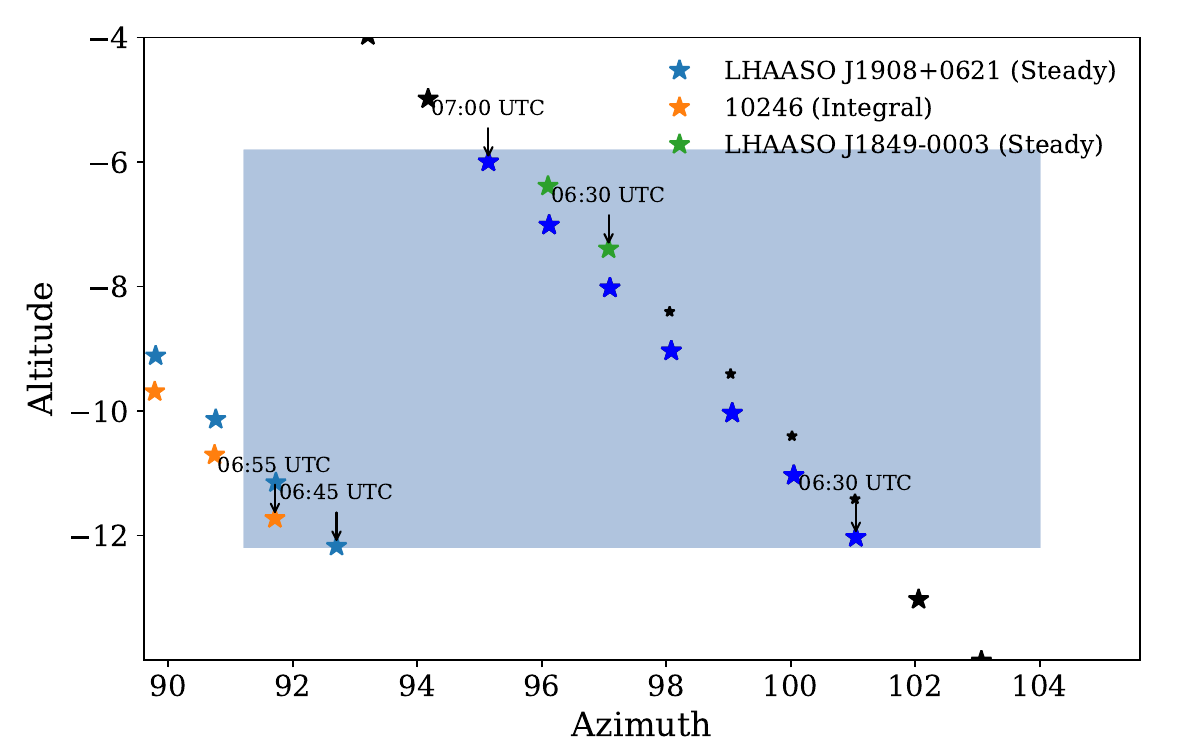}
    \caption{Trajectories of two sources scheduled for observation (blue stars) for similar conditions as figure~\ref{fig:DetectorTrajectory}, in the balloon frame as a function of azimuth and altitude, respectively TDE 2022lri (left panel) and Cascade 137930\_36941090 (right panel). Superimposed is the detector field of view (blue rectangle). We also show other sources that cross the field of view during this observation period (colored stars and source names in the legend).}\label{fig:Trajectories}
\end{figure}

The scheduler module output provides the user with a list of sources with their pointing times and directions (in the detector frame: azimuth and altitude) that can be used by telescope operators or burst advocates. As discussed in section~\ref{sec:sunmoon}, for the Sun and Moon observing conditions and a mock source database built from GCN, TNS and ATels alerts as well as a steady source catalog, observable sources for one day are shown in figure~\ref{fig:SkyMap} (lower panel). The observation window, limited by the Sun and Moon illumination, only permits three scheduled sources, represented by colored stars. Two source trajectories in the detector's frame are illustrated in figure~\ref{fig:Trajectories}, respectively TDE 2022lri and Cascade 137930\_36941090. The first one is a tidal disruption event (TNS alert) detected on 2022-06-01 by ATLAS \citep{2022TNSTR1521....1T} and the second one is a neutrino event (GCN alert) detected on 2023-05-11 by IceCube. In this figure, coincident observations are also shown to highlight the risk of confusion with other sources.

\subsubsection{Adapting the scheduler module to other detectors}

Several aspects of the scheduler module are specific to the EUSO-SPB2 and PBR missions, but can be easily adapted to other types of detectors requiring pointing for their observing programs. The parameters determining the scheduling of successive observations, such as the FOV extension in altitude and azimuth, but also the time required for re-pointing, the minimum observation time for one source and the maximum number of re-pointings, can be changed in the input parameters depending on the requirements of specific missions. More complex FOV configurations, for instance depending on topography, and different observation strategies, for instance determined by the detector's acceptance as a function of the energy, require further developments.

\subsection{Visualizations}

Every module includes a set of optional visualizations, that include:
\begin{itemize}
    \item sky maps of the payload trajectory and the observation window as illustrated in figure~\ref{fig:DetectorTrajectory},
    \item sky maps in Galactic or equatorial coordinates showing the observable portion of the sky and potentially overlaying all or a subset of sources from the source catalog (for instance only observable sources, or only scheduled sources) as illustrated in figure~\ref{fig:SkyMap},
    \item weighted time as a function of azimuthal pointing for given gravitational wave events, as illustrated in  figure~\ref{fig:GWresult},
    \item trajectories of sources in the payload frame superimposed with the detector FOV and coincident observations, as illustrated in figure~\ref{fig:Trajectories},
    \item successive observation windows for a full flight path overlapped with Sun and Moon observability constraints, as illustrated in figure~\ref{fig:Obs_window_SB},
    \item cumulated observation times for observable and scheduled sources for a full flight path, for each priority ranking or for all sources, as illustrated in figures~\ref{fig:Obs_time_SB} and \ref{fig:Obs_time_SB_strat34}.
\end{itemize}
To create a specific figure when running NuTS, its name needs to be given in the configuration file. However, figure names are not initialized by default in the configuration file and need to be added by the user. The full list of default figure names is given in the online documentation.

\subsection{Performance}

We optimized NuTS to enable rapid evaluation of source observability and efficient production of observation schedules. These capabilities are crucial during observation campaigns, both for preparing schedules in advance and for quick re-scheduling, when necessary.

For the point source algorithm, we evaluated the runtime using a time interval of $10$~min and a catalog with a varying number of sources. The observability module’s runtime ranges from 1 to 5 seconds (typically accounting for 30\% of the total computing time), while the scheduling module takes between 1 and 15 seconds (typically 70\% of the computing time). Both runtimes scale linearly with the number of sources in the catalog. The visualization routines, however, are more resource-intensive: for example, plotting all observable source trajectories crossing the detector’s field of view (via the \verb|source_trajectories_zoom| option in the configuration file) doubles the runtime.

For the poorly localized sources algorithm, the number of effective point sources and the time interval of $1$~min required to determine observability increase the runtimes. For our example of S190425z,
the 90$\%$ region has a total of 12,132 points in the localization region with a total runtime of $\sim 11$ minutes, while the event S240422ed has a runtime of $\sim 18$ seconds with a smaller localization region represented by 482 points. 

\section{Scheduling observations}\label{sec:schedules}

In this section, we provide a couple of concrete examples of schedules that can be produced by our software, with a focus on the specific properties of the EUSO-SPB2 mission. To do so, we compute several sets of mock schedules between 2023-04-16 and 2023-05-24 using a source database built from GCN, TNS, and ATels alerts as well as a steady source catalog. In addition, we use the flight data of the SuperBit mission \citep{Gill:2024mqb}, which flew between 2023-04-16 and 2023-05-24, giving us access to a realistic balloon trajectory.

\begin{figure}[ht]
    \centering
    \includegraphics[width=.49\textwidth]{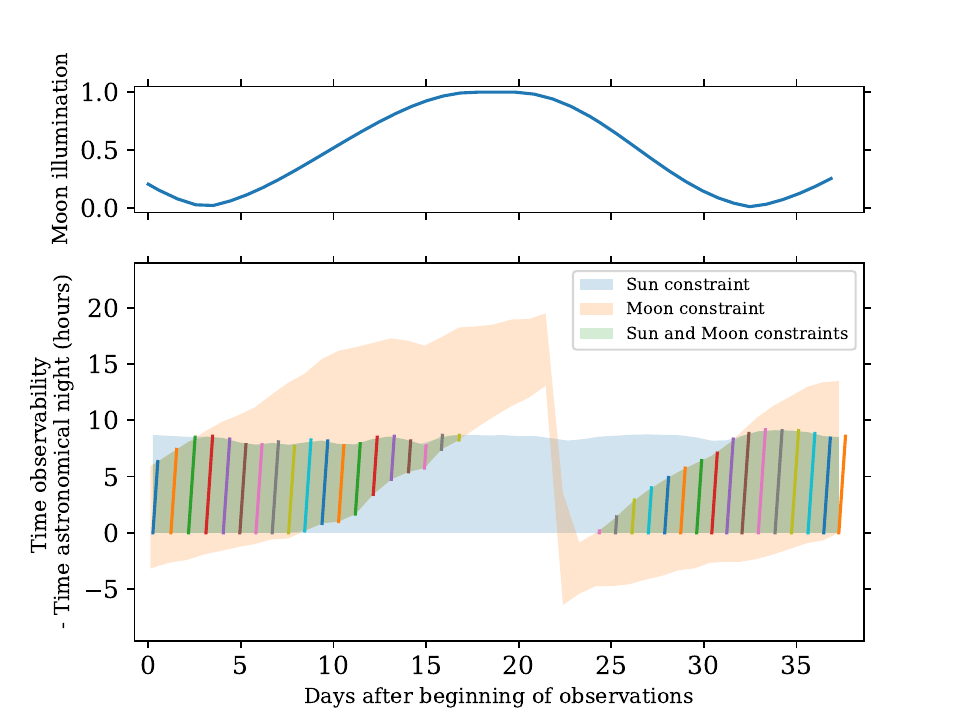}
    \caption{Moon illumination (top panel) and successive observation windows (bottom panel) as a function of days after the beginning of observations for the SuperBIT flight path between 04-16-23 and 05-24-23. We superimpose the times of observability using the Sun constraint, the Moon constraint, and the Sun and Moon constraints (see legend). The successive observation windows are represented by the colored lines.}
    \label{fig:Obs_window_SB}
\end{figure}

In figure~\ref{fig:Obs_window_SB}, we show the successive observation windows that we obtain between those dates, using the run option \verb|obs_windows_all| of NuTS. Observations are possible when the Sun constraint band (blue, astronomical night) and Moon set/low illumination (orange region) overlap. No observations can be scheduled between 2023-05-02 and 2023-05-11 due to the joint constraints from the Sun and the Moon, in particular due to the Moon illumination close to full Moon. In addition, this run option allows us to calculate some general characteristics of this mock observation campaign. The total flight time of the SuperBit mission is $\sim 3 \times 10^6\, {\rm s}$, and during this flight the cumulated duration of all observation windows is $8 \times 10^5\,{\rm s}$. Thus, observations can be scheduled during $23\%$ of the flight time.

In addition, NuTS allows comparisons of the properties of the sources scheduled for several scheduling strategies. As an example, in this section we compare two scheduling strategies, hereinafter referred to as 3 and 4. For the scheduling strategy 3, to select the sources we use the \textit{source type criterion} with \verb|strategy_priority_sort = True| and \verb|strategy_priority_max = 3|. Thus, we select all sources that can be scheduled up to priority ranking $<3$, and then sample one of each priority ranking until the schedule is filled. We also use the \textit{observation time criterion}, with \verb|strategy_obstime = True|. Therefore, for each ranking we try to schedule the sources that can be observed for the longest time first. For the scheduling strategy 4, the strategy is similar, however we add the \textit{previous observation criterion} with with \verb|strategy_obsprev = True|, thus the algorithm attempts first to schedule the sources that have been scheduled during the previous observation window. This strategy aims at maximizing the total observation time for several interesting sources, while strategy 3 allows a priori observation of a wider variety of sources.

\begin{figure}[ht]
    \includegraphics[width=0.48\textwidth]{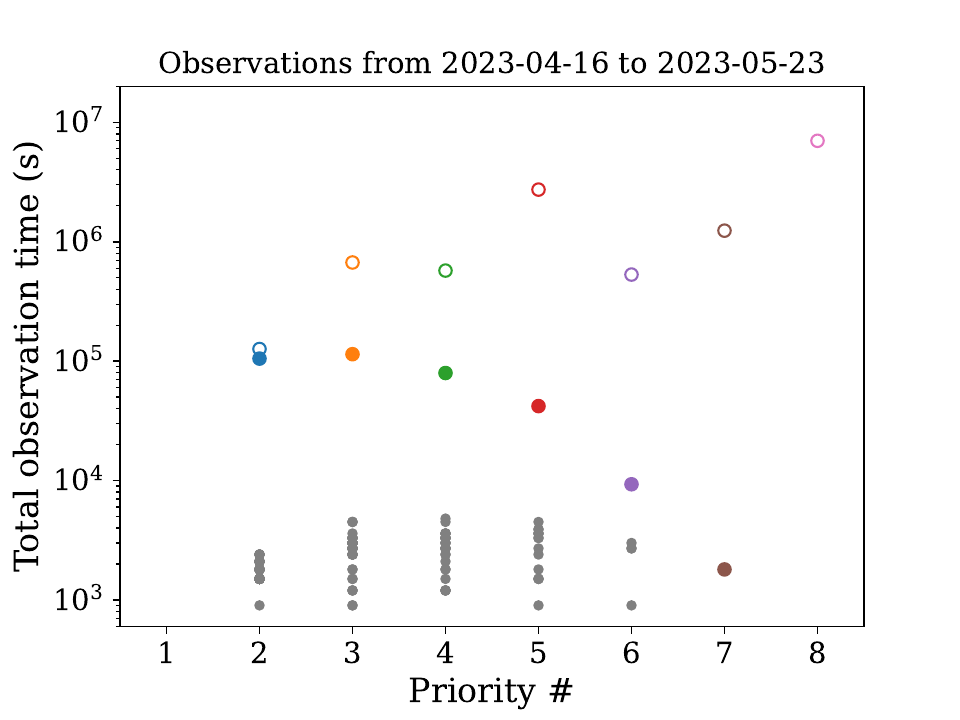}
    \includegraphics[width=0.48\textwidth]{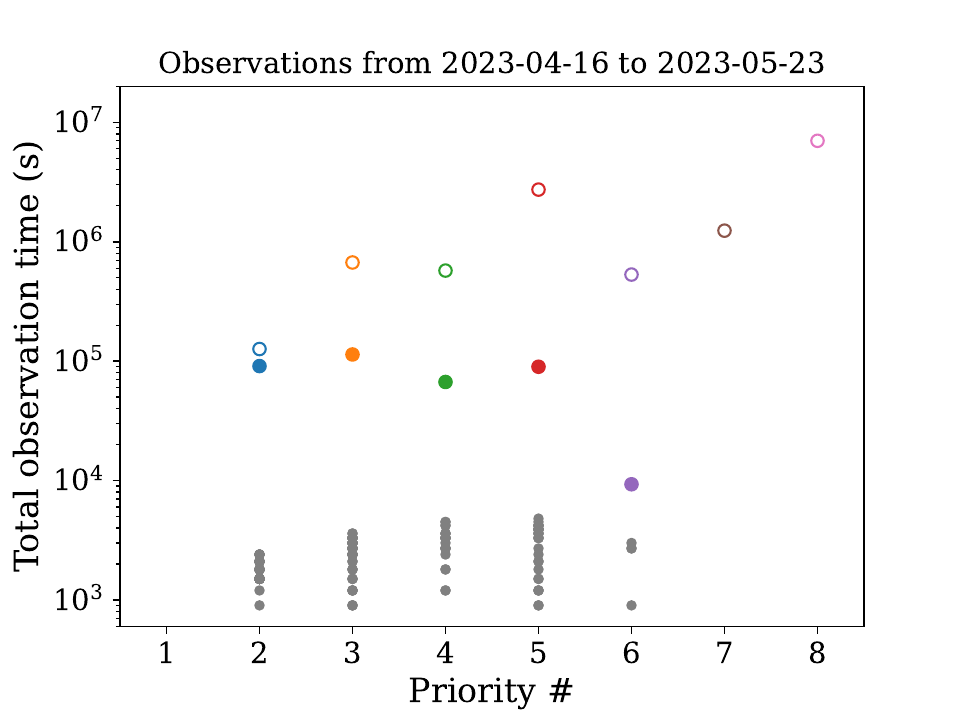}
    \caption{Total observation times (s) for scheduled sources (colored points) and observable sources (colored circles), in log scale, for the eight priority classes, using the trajectory of the SuperBit flight. We also show the observation times for individual observation windows (grey points). We compare the results obtained for the scheduling strategy 3 (left panel) and the scheduling strategy 4 (right panel). See text for more detail.}
    \label{fig:Obs_time_SB}
\end{figure}

We show the results in figure~\ref{fig:Obs_time_SB}, where we compare the total observation follow-up time of various sources selected while running our mock scheduling, for the scheduling strategies 3 and 4. More precisely, we show the total observation times for scheduled sources and observable sources, summed for each priority ranking. The details of the observation schedules can be found in \ref{app:app_schedules}. Sources between priority ranking $2$ and $6$ are scheduled. A wider variety of sources are scheduled for strategy 3. The cumulated observation times of individual sources range from $\sim 9 \times 10^2\,{\rm s}$ to $\sim 7 \times 10^4\,{\rm s}$, and the total cumulated observation time is $\sim 4 \times 10^5\,{\rm s}$, about half of the cumulated duration of all observation windows, and thus $\sim 11\%$ of the total flight time.

\section{Summary and Prospects}\label{sec:prospects}

NuTS is a python package that was developed by the JEM-EUSO collaboration, in the context of the EUSO-SPB2 mission and its search of VHE neutrinos associated with energetic transient and steady astrophysical sources. This software is modular, comprised of a listener module, an observability module and a scheduler module. NuTS can produce databases of transient and steady sources, by selecting prospective sources of VHE neutrinos. In addition, it can compute the list of observable sources for given observation windows and a variety of observability criteria. Finally, it can produce observation schedules following different prioritization criteria. This software can be used in real time (with a full run time of only a few minutes for one observation window), or to produce mock lists of observable sources and observation schedules in order to evaluate the performances of a prospective mission.

In the near future, a second version NuTS tailored to properties of the PBR mission \citep{Battisti:2024jjy} will be developed. During the PBR mission, scheduled for a launch in 2028, NuTS will be used to produce custom observation schedules for VHE neutrino searches from transient sources. In this context, several important developments will be required, including an update of the listener module and a coupling with the JEM-EUSO collaboration tools used for sensitivity calculations.

First, the listener module will be updated in order to adapt to the rapidly evolving alert systems. When possible, our developments will take advantage of the existing collaborative tools focusing on alert collection and distribution for the observation of explosive transients (for instance, the Astro-COLIBRI platform  \citep{Reichherzer:2021pfe}). New types of alerts will be included. The large number of alerts from VRO will provide insight into many transient categories, e.g., TDEs or a great variety of cc-SNe classes. The alerts and observations from new missions such as SVOM and Einstein Probe will provide insight into interesting GRB events. In addition, we will work on automating alert collection when machine-readable formats are not available, in particular from ATels.

Furthermore, NuTS will be coupled with $\nu$SpaceSim in order to have access not only the observation times of the scheduled sources but also to their acceptance. This development will allow to directly couple NuTS to sensitivity computations. Therefore, in the case of the absence of VHE neutrino observations during transient source follow-ups, our tools will be able to perform rapid calculations of sensitivity limits, that are useful to guide multi-messenger follow-up observations.

In the longer term, we will adapt NuTS to satellite trajectories (ex. for the prospective POEMMA mission \citealp{POEMMA:2020ykm}) and ground detectors (e.g. GRAND \citealp{GRAND:2018iaj}, BEACON \citealp{Wissel:2020sec} or HERON \citealp{GRAND:2025rps}) in order to make it available for the various types of detectors focusing on VHE neutrino searches from energetic transients, which will allow precise comparisons between detection techniques. Developers interested in specific features can access the source code on gitlab. They can submit tickets if they encounter some issues, and are encouraged to contact the development team for guidance and collaboration on new developments.

\noindent{\bf Acknowledgements}\\
This work was supported by NASA awards 80NSSC22K1488, 80NSSC18K047, and the French space agency CNES.
\appendix

\section{Documentation}\label{appendix:documentation}
The documentation for this package is hosted on gitlab and can be found at this link \url{https://jem-euso.gitlab.io/euso-spb2/too/too-nuts/}. It can also be built by the user using Sphinx. First, install Sphinx with pip. Then run the following commands:
\begin{lstlisting}
pip install .
pip install -r docs/requirements
mkdir docs_b
sphinx-build -b html docs docs_b
\end{lstlisting}
The documentation is then available in the directory \verb|docs_b| in html format.

\section{Installation}\label{app:intallation}
\subsection{Installing from pip}
The information about the NuTS package can be found at this link \url{https://pypi.org/project/too-nuts/}. NuTS can be installed via pip using
\begin{lstlisting}
pip install too-nuts
\end{lstlisting}

\subsection{Installing from source}
Download the repository from gitlab (\url{https://gitlab.com/jem-euso/euso-spb2/too/too-nuts.git}). Make sure pip and python (version $\geq 3.10$) are available. In the source directory
\verb|too-nuts| install with
\begin{lstlisting}
pip install .
\end{lstlisting}
Requirements are defined in the file \verb|setup.cfg|. Missing Python packages should be downloaded and installed automatically during the installation.

\subsection{Alert systems}\label{appendix:alert}

NuTS includes an interface to two main transient alert networks, TNS and GCN, that can be monitored in near real-time. For this purpose, the listener modules have been developed, which can be run using
\begin{lstlisting}
nuts listen <config> -l [GCN | TNS]
\end{lstlisting}

\subsubsection{TNS}
The TNS listening module downloads an updated database from TNS at a predefined cadence (typically 1h and we do not recommend downloading at a too high cadence). To be able to use this service, the user has to create a user account on the TNS webpage\footnote{See the TNS getting started page: \href{https://www.wis-tns.org/content/tns-getting-started}{https://www.wis-tns.org/content/tns-getting-started}.} and replace the following fields in the configuration file.
\begin{lstlisting}
[settings.TNS]
user_id = <user_id>
user_name = <user_name>
\end{lstlisting}
The other arguments in the \verb|[settings.TNS]| tag can be used to specify the output format and download frequency.
\subsubsection{GCN}
GCN sends out real-time alerts via its GCN-Kafka network. First, the user has to register to receive the GCN notices\footnote{See the GCN getting started page: \href{https://gcn.nasa.gov/quickstart}{https://gcn.nasa.gov/quickstart}.} and replace the following fields in the configuration file:

\begin{lstlisting}
[settings.GCN]
client_id = <client_id>
client_secret_name = <client_secret>
\end{lstlisting}

The GCN alerts are received by the GCN listener module, which extracts the primary information from the alert and creates an event following the data format used in NuTS. There are different alert-type formats for different instruments and different processing stages. We developed a template scheme for many of these alerts, and will expand and complete this list for the next NuTS release. A list of all the supported alerts can be found in the software source directory, in the file \verb|GCN_alerts.csv|. In the same file, the user can toggle the parsing on or off for any of these alerts using a boolean, depending on the unique requirements of the experiment. 

\subsubsection{Other alerts}
For all other alerts, for instance alerts displayed in ATels, no dedicated pipeline has been developed yet. However, they can easily be added manually into a csv. The name of this cvs file (by default ``OtherTransients.csv'') should be provided in the configuration file
\begin{lstlisting}
[files.database]
other = "OtherTransients.csv"
\end{lstlisting}
This csv must contain the keys required by the data format used in NuTS (see \ref{app:data-format}) and can be used as part of the database.

\section{Usage}\label{app:usage}

\subsection{Command line interface}

The command line interface (CLI) for the Neutrino Target Scheduler is based on clicker and works as follows
\begin{lstlisting}[language=bash]
nuts COMMAND [OPTIONS] [ARGS]
\end{lstlisting}
with several available commands: \verb|init|, \verb|make-config|, \verb|listen|, \verb|run|, \verb|single| and \verb|gui|. For nuts and all commands, information is accessible with the option \verb|--help|.

The \verb|init| command is used to set up the directory structure for NuTS. This includes creating the necessary folders and configuration files used for the source database. NuTS by default includes a short list of hand selected source candidates. This is used as follows:
\begin{lstlisting}[language=bash]
nuts init /path/to/your/Database
\end{lstlisting}

The \verb|make-config| command can be used to generate a configuration file with the correct path to the installed version of nuts, and is used as follows:
\begin{lstlisting}[language=bash]
nuts make-config <config-file-name.toml>
\end{lstlisting}
We recommend removing all the parameters that the user does not want to edit. The input values that are not provided will be filled with the default values. The names of the figures are not automatically added to the configuration file, and figures are not created when their names are not written in this file. More information is provided in the documentation.

The \verb|listen| command is dedicated to the listeners for GCN and TNS. For example 
\begin{lstlisting}[language=bash]
nuts listen <config-file-name.toml> -l GCN
\end{lstlisting}
starts the GCN listener. Received alerts are saved in the Catalogs directory and a log file is created.

The \verb|run| command is the main command used to trigger the scheduler. For example
\begin{lstlisting}[language=bash]
nuts run <config-file-name.toml> -o <option>
\end{lstlisting}
with \verb|<option>| set to \verb|all| prepares the database, calculates a possible observation window, finds observable sources and builds a schedule. Run currently allows for the following options:
\begin{itemize}[noitemsep,topsep=0pt]

   \item \verb|obs_window|: calculates the next possible observation window(s) for the input date given in the configuration file

   \item \verb|combine_db|: prepares the database by combining the different database files

   \item \verb|clean_db|: prepares the database by excluding outdated sources and adding the priority ranking for the sources

    \item \verb|prep_db|: \verb|combine_db| and \verb|clean_db|

    \item \verb|observability|: calculates the observable sources for a given database

    \item \verb|observations|: \verb|prep_db| and \verb|observability|

    \item \verb|schedule|: calculates a schedule for a list of known observable sources

    \item \verb|obs_sched|: \verb|observability| and \verb|schedule|

    \item \verb|gw|: runs observability for poorly localized (GW) sources

    \item \verb|pointing_obs|: computes FOV cuts for a known pointing of the detector

    \item \verb|visuals|: produces visualizations of the results

    \item \verb|all|: \verb|prep_db|, \verb|obs_sched| and \verb|visuals|

    \item \verb|obs_windows_all|: computes all successive observation windows for a given flight time and trajectory

    \item \verb|flight|: computes all observabilities, schedules and visuals for a given flight time and trajectory
\end{itemize}

The \verb|single| command allows to compute the observability of a single source. All the information about the source need to be specified as input options, for instance
\begin{lstlisting}[language=bash]
nuts single <config-file-name.toml> -ra <right ascension> -dec <declination> -dt <detection time> ...
\end{lstlisting}
All the input options are explained in \verb|nuts single --help|.

\subsection{Graphical user interface}
A graphical user interface (GUI) allows the user to perform most of the actions allowed by the CLI. To start this interface, use the command line
\begin{lstlisting}[language=bash]
nuts gui
\end{lstlisting}
The GUI provides a documentation, allows one to generate and edit a configuration file, listen to alert systems, add sources to the database, run NuTS to determine observable source and compute an observation schedule, schedule a single source, visualize the results. The ToO user interface was developed using the open-source Python framework Streamlit. 

\subsection{Usage for developers}
The script used by the command line interface, namely \verb|compute.py|, can be an example for developers who want to create their own features. This script is located in the source directory. All primary use case-dependent pieces, such as the detector location, sun, moon, field of view cuts, and scheduler strategies, follow a custom interface structure to make it easy to expand and adapt for other experimental needs. 

In general, after installation, classes and functions are programmatic accessible, as modules are independently importable sub-packages under nuts.*. Each module takes typed inputs (dataclasses / pydantic models) and returns typed outputs, so you can use them freely after importing them in your user code. Here is a minimal example to calculate the moon illumination and phase angle:

\begin{lstlisting}[language=Python, caption=Programmatic access example]
from astropy.time import Time
from nuts.observation_period.sun_moon_cuts import moon_phase_angle, moon_illumination

# New moon (Oct 2024): phase angle near 180, illumination near 0%
new_moon = Time("2024-10-02T18:00:00", format="isot", scale="utc")
print(f"Phase angle:  {moon_phase_angle(new_moon).to(u.deg):.1f}")
print(f"Illumination: {moon_illumination(new_moon):.1%}")

# Full moon (Oct 2024): phase angle near 0, illumination near 100
full_moon = Time("2024-10-17T11:00:00", format="isot", scale="utc")
print(f"Phase angle:  {moon_phase_angle(full_moon).to(u.deg):.1f}")
print(f"Illumination: {moon_illumination(full_moon):.1%}")
\end{lstlisting}

\section{Data format}\label{app:data-format}

Events associated with transient or steady sources contain the following information:
\begin{itemize}[noitemsep,topsep=0pt]
    \item \verb|event_type|: str (``GRB", ``TDE", ...)
    \item \verb|event_id|: str (``GRB123456", ...)
    \item \verb|publisher|: str (``Fermi'', ``Swift'', ``TNS'', ...)
    \item \verb|publisher_id|: str (``ATels123456")
    \item \verb|coordinates|: astropy.SkyCoord (ra, dec)
    \item \verb|detection_time|: atime.Time (``2022-11-11T11:11:11")
    \item \verb|params|: dict (any parameters that might be interesting for the event)
\end{itemize}

Several units conventions are adopted in the software, in general by using quantities with their associated units defined by astropy, for instance for times, coordinates and angles.

\section{Field testing with starlight}\label{app:test}

\begin{figure}[ht]
    \centering
    \includegraphics[width=0.49\textwidth]{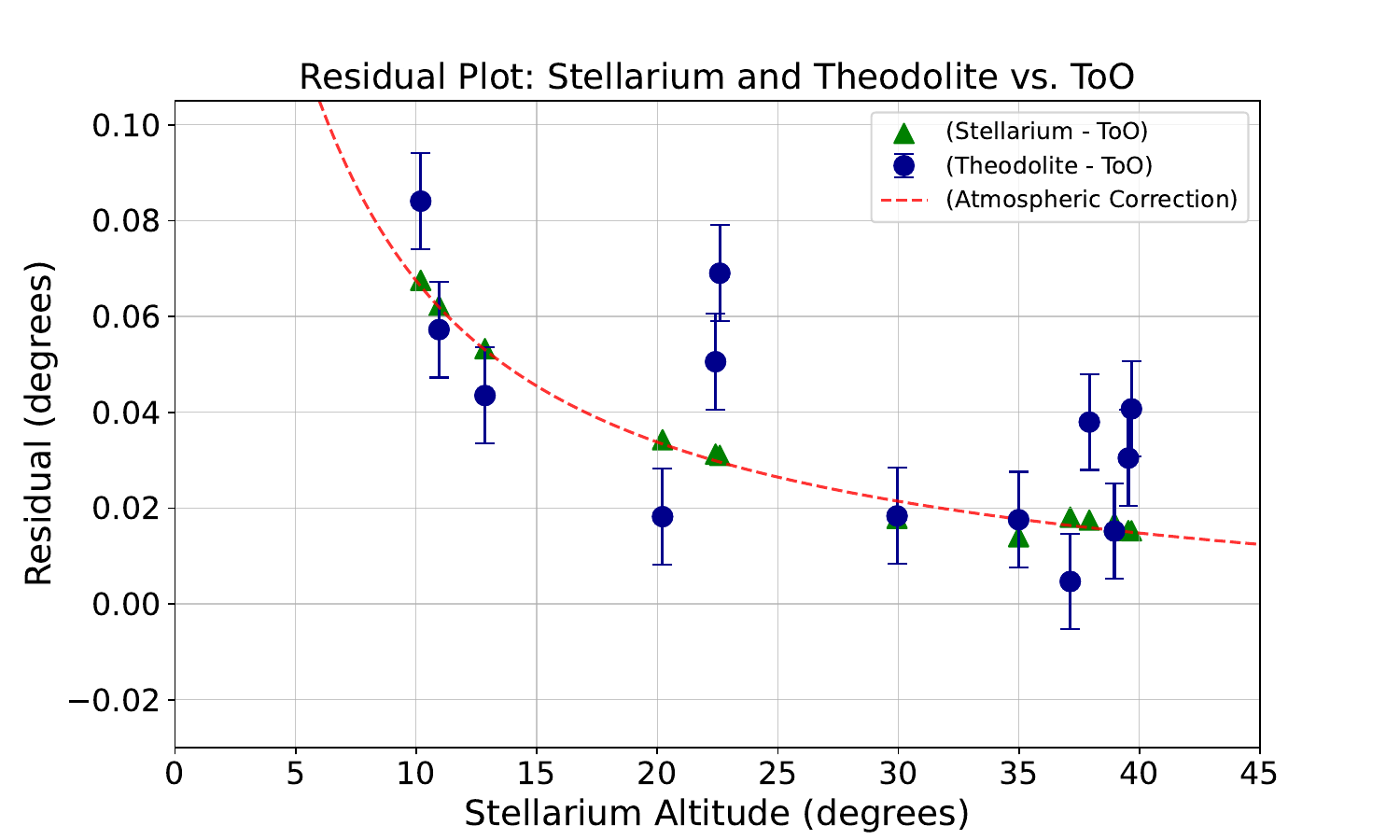}
    \includegraphics[width=0.49\textwidth]{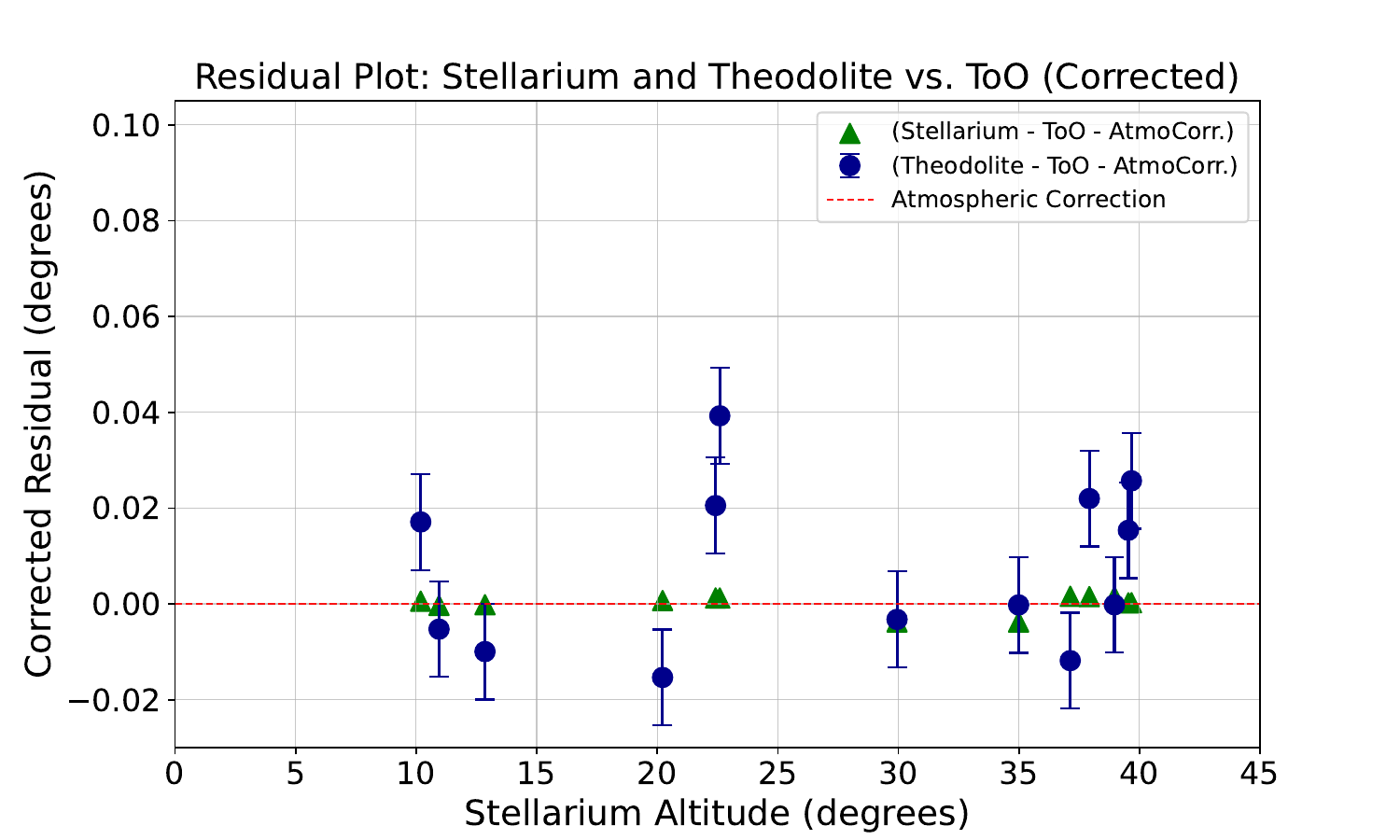}
    \caption{Left panel: differences between the measured theodolite altitude of visible stars and their altitudes determined by the NuTS (dots) ToO scheduling software and the difference between the altitudes as predicted by Stellarium and by NuTS (triangles) as a function of (Stellarium) altitude. The dashed line shows the atmospheric correction for observations at 2300 m elevation where the tests were done. Right panel: same as left panel, after the atmospheric correction is applied.}
    \label{fig:ToO_Residual}
\end{figure}

The NuTS software is applicable for scheduling neutrino ToO source observations as well as observations of other astrophysical objects, both above and below the limb. A test of the NuTS software is described here. A recently calibrated Topcon DT-307L theodolite was used to measure the angle above the horizontal direction (altitude) of Arcturus, Sirius, Rigel, Betelgeuse, and Altair from a dark sky site at $2300$\,m elevation near the Great Sand Dunes National Park in Colorado USA during a moonless night. The UTC time of each measurement was also recorded. The predicted altitude angles of the measurements were generated by the NuTS software for the same stars using the location, elevation, and times of the theodolite measurements. As an additional cross-check, the Stellarium software \citep{Zotti_Hoffmann_Wolf_Chereau_Chereau_2021, Zotti_Wolf_2022} was also used to determine the altitude angles. Figure \ref{fig:ToO_Residual} shows the difference between the measured theodolite altitudes of the stars and their altitude determined in NuTS (dots) and the difference between the Stellarium and NuTS altitudes (diamonds). These residuals (in degrees) are less than 0.1$^\circ$ for altitudes measured in the range of $10^\circ-40^\circ$. The estimated error bars on the theodolite measurement points include uncertainties arising from the leveling of the theodolite, the centering of the theodolite's altitude direction on a moving star, and the timing precision.

The residuals are not closer to $0^\circ$ because of light refraction in the atmosphere. The dashed line in figure~\ref{fig:ToO_Residual} shows the correction due to refraction. The bottom panel of figure~\ref{fig:ToO_Residual} shows the residuals accounting for the correction for refraction. Corrected residuals are less than $\sim 0.05^\circ$ for the range of altitudes measured in the test. Since the theodolite measured the observed direction of the stars using starlight, atmospheric refraction effects are inherently present. The Stellarium software as used for this test included atmospheric refraction. The NuTS software does not have atmospheric refraction effects included, as it is designed primarily to schedule observations of neutrino sources through the detection of extensive air showers.

\section{Detailed observation schedules}\label{app:app_schedules}

In figure~\ref{fig:Obs_time_SB_strat34}, we show the detail of figure~\ref{fig:Obs_time_SB}, i.e. the total observation times (s) in log scale for all the sources that could have been scheduled during the trajectory of the SuperBit flight using our mock catalog of sources. The observation strategies 3 and 4 are compared (see main text for detail). In addition we show examples of observation schedules for two selected dates. These schedules provide information about the source type, the pointing time (UTC), the times of beginning and end of observation (UTC), and the pointing direction in altitude and azimuth (in degrees). A maximum delay of $10\,{\rm min}$ (given by the parameter \verb|rotation_time|) is given for the pointing in azimuth. The pointing in zenith does not change as the detector always points below the Earth's limb. Sources scheduled during one observation window tend to be rescheduled during the next observation window, due to their sky location; this effect is enhanced with the scheduling strategy 4.

\begin{figure*}[p]
    \includegraphics[width=0.98\textwidth]{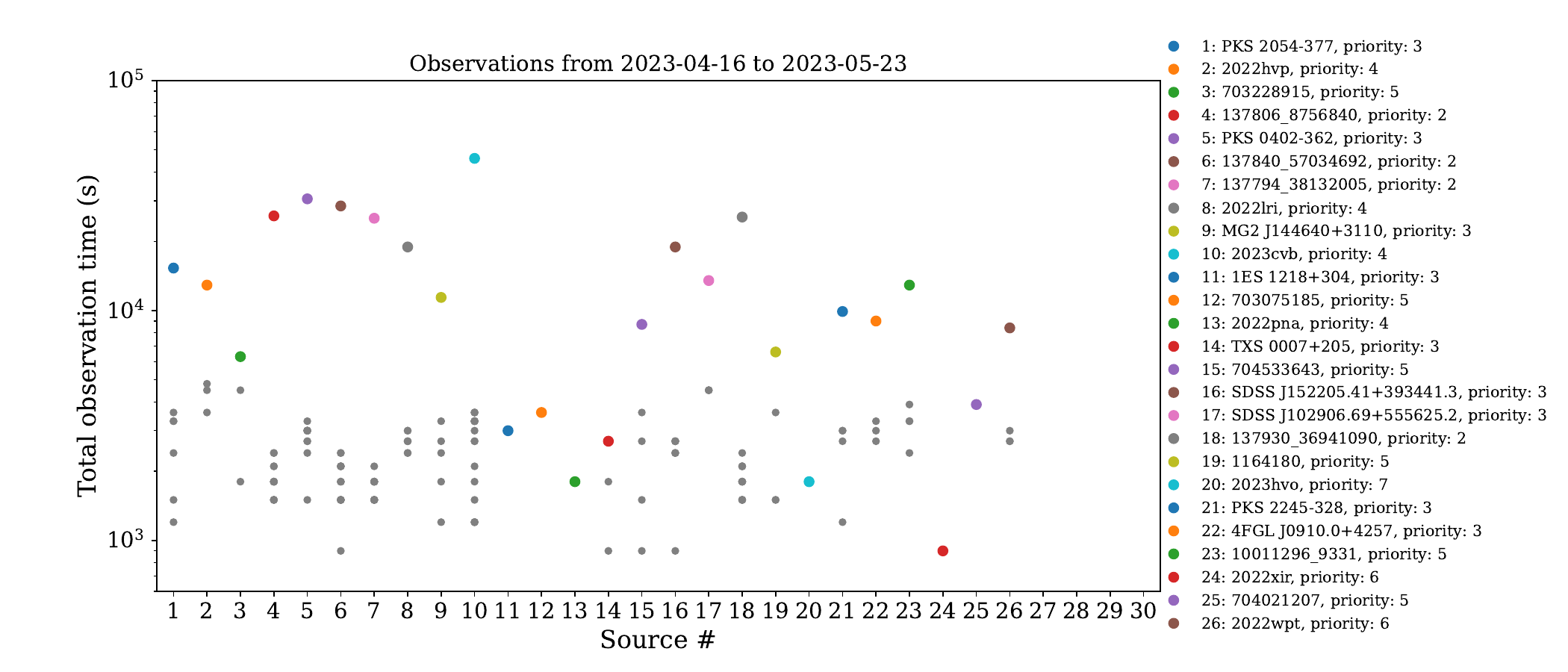}\\
    \includegraphics[width=0.98\textwidth]{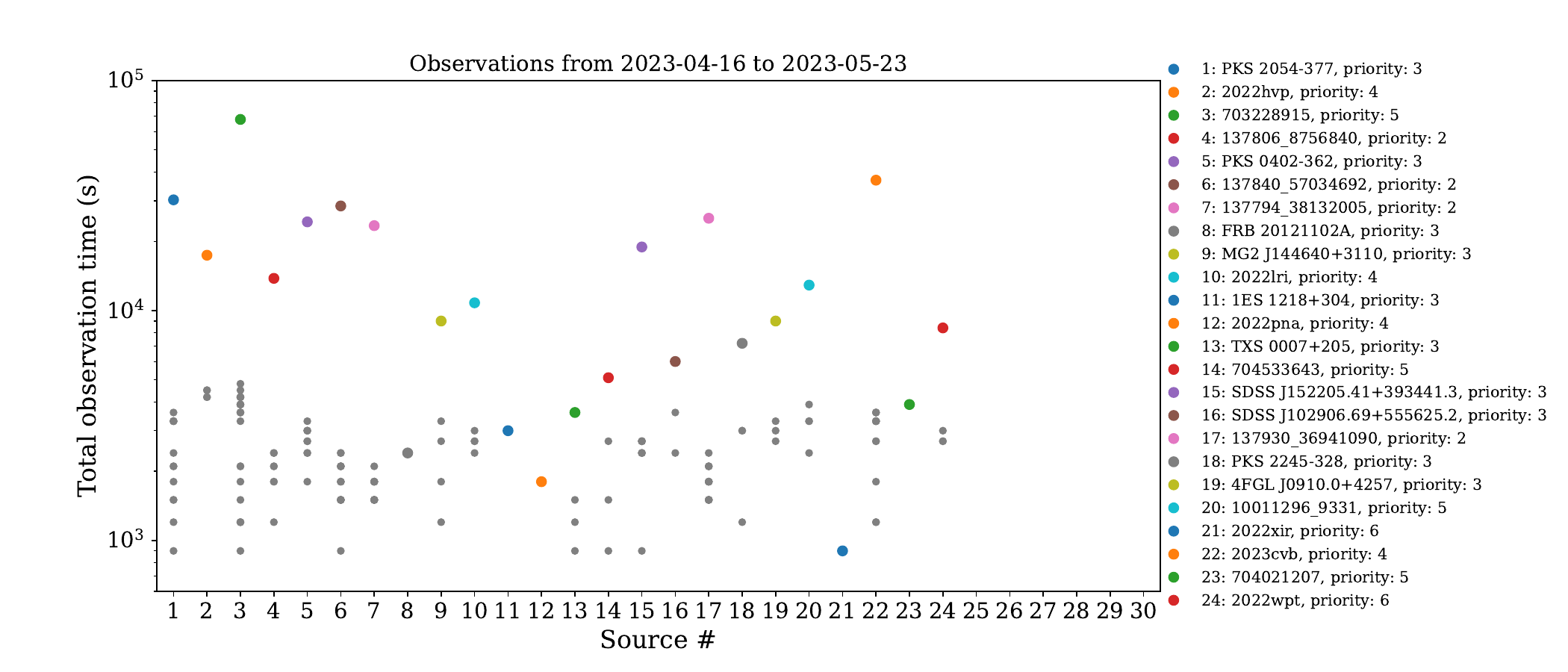}
\end{figure*}

\begin{figure*}[p]
    {\scriptsize
        \textbf{2023-04-16 - Schedule of observations}\\
        \begin{tabular}{p{0.3\textwidth} p{0.3\textwidth} p{0.3\textwidth}}\hline
        \textbf{Source observed} & \textbf{Pointing and observation times (UTC)} & \textbf{Pointing direction (altitude, azimuth)}\\\hline
        PKS 2054-377 (FSRQ) & 07:55, [08:05 - 09:05] & (-9.01\degree,  171.40\degree)\\\hline
        PKS 2054-377 (FSRQ) & 09:05, [09:15 - 09:35] & (-9.01\degree, 162.01\degree)\\\hline
        2022hvp (TDE) & 09:35, [09:45 - 11:00] & (-9.00\degree, 347.46\degree)\\\hline
        703228915 (GRB) & 11:05, [11:15 - 12:30] & (-9.01\degree, 23.69\degree)\\\hline
        137806\_8756840 (IceCube BRONZE) & 12:30, [12:40 - 13:10] & (-9.01\degree, 276.25\degree)\\\hline
        \end{tabular}\vspace{0.2cm}\\
        \textbf{2023-04-17 - Schedule of observations - strategy 3}\\
        \begin{tabular}{p{0.3\textwidth} p{0.3\textwidth} p{0.3\textwidth}}\hline
        \textbf{Source observed} & \textbf{Pointing and observation times (UTC)} & \textbf{Pointing direction (altitude, azimuth)}\\\hline
        2022hvp (TDE) & 07:15, [07:25 - 08:45] & (-9.01\degree, 359.58\degree) \\\hline
        2022hvp (TDE) & 08:45, [08:55 - 09:55] & (-9.01\degree, 348.70\degree) \\\hline
        137806\_8756840 (IceCube BRONZE) & 11:40, [11:50 - 12:20] & (-9.01\degree, 275.91\degree) \\\hline
        PKS 0402-362 (blazar) & 12:20, [12:30 - 13:15] & (-9.00\degree, 194.92\degree) \\\hline
        PKS 0402-362 (blazar) & 13:15, [13:25 - 14:15] & (-9.01\degree, 182.31\degree) \\\hline
        \end{tabular}\vspace{0.2cm}\\
        \textbf{2023-04-17 - Schedule of observations - strategy 4}\\
        \begin{tabular}{p{0.3\textwidth} p{0.3\textwidth} p{0.3\textwidth}}\hline
        \textbf{Source observed} & \textbf{Pointing and observation times (UTC)} & \textbf{Pointing direction (altitude, azimuth)}\\\hline
        PKS 2054-377 (FSRQ) & 07:15, [07:25 - 08:20] & (-9.01\degree, 171.66\degree) \\\hline
        2022hvp (TDE) & 08:20, [08:30 - 09:40] & (-9.01\degree, 352.06\degree) \\\hline
        703228915 (GRB) & 10:25, [10:35 - 11:55] & (-9.00\degree, 21.86\degree) \\\hline
        PKS 0402-362 (blazar) & 11:55, [12:05 - 13:00] & (-9.01\degree, 198.51\degree) \\\hline
        PKS 0402-362 (blazar) & 13:00, [13:10 - 14:00] & (-9.01\degree, 185.49\degree) \\\hline
        \end{tabular}
        }
    \caption{Total observation time (s) in log scale for the sources that could have been scheduled using the trajectory of the SuperBit flight (colored points). We also show the observation times for individual observation windows (grey points). We show the results obtained for the scheduling strategy 3 (top panel) and strategy 4 (middle panel) and three examples of schedules (bottom panel). See text for more detail.}
    \label{fig:Obs_time_SB_strat34}
\end{figure*}

{\small
\bibliographystyle{elsarticle-num-names} 
\bibliography{ToO}

@article{Eser:2023lck,
    author = "Eser, Johannes and Olinto, Angela V. and Wiencke, Lawrence",
    collaboration = "JEM-EUSO",
    title = "{Overview and First Results of EUSO-SPB2}",
    eprint = "2308.15693",
    archivePrefix = "arXiv",
    primaryClass = "astro-ph.HE",
    doi = "10.22323/1.444.0397",
    journal = "PoS",
    volume = "ICRC2023",
    pages = "397",
    year = "2023"
}

@article{Abe:2023lne,
    author = "Abe, S. and others",
    title = "{JEM-EUSO Collaboration contributions to the 38th International Cosmic Ray Conference}",
    eprint = "2312.08204",
    archivePrefix = "arXiv",
    primaryClass = "astro-ph.HE",
    month = "12",
    year = "2023"
}

@article{Chu:2015jxa,
    author = {Chu, Q. and Howell, E. J. and Rowlinson, A. and Gao, H. and Zhang, B. and Tingay, S. J. and Bo{\"e}r, M. and Wen, L.},
    title = "{Capturing the electromagnetic counterparts of binary neutron star mergers through low latency gravitational wave triggers}",
    eprint = "1509.06876",
    archivePrefix = "arXiv",
    primaryClass = "astro-ph.HE",
    doi = "10.1093/mnras/stw576",
    journal = "Mon. Not. Roy. Astron. Soc.",
    volume = "459",
    number = "1",
    pages = "121--139",
    year = "2016"
}

@article{JEM-EUSO:2023ypf,
    author = "Abdellaoui, G. and others",
    collaboration = "JEM-EUSO",
    title = "{EUSO-SPB1 mission and science}",
    eprint = "2401.06525",
    archivePrefix = "arXiv",
    primaryClass = "astro-ph.IM",
    doi = "10.1016/j.astropartphys.2023.102891",
    journal = "Astropart. Phys.",
    volume = "154",
    pages = "102891",
    year = "2024"
}

@article{Adams:2024gsj,
    author = "Adams, Jr., James H. and others",
    title = "{The EUSO-SPB2 fluorescence telescope for the detection of Ultra-High Energy Cosmic Rays}",
    eprint = "2406.13673",
    archivePrefix = "arXiv",
    primaryClass = "astro-ph.IM",
    doi = "10.1016/j.astropartphys.2024.103046",
    journal = "Astropart. Phys.",
    volume = "165",
    pages = "103046",
    year = "2025"
}

@article{Cummings:2023ypo,
    author = "Cummings, Austin",
    collaboration = "JEM-EUSO",
    title = "{Analysis of above-the-limb cosmic rays for EUSO-SPB2}",
    eprint = "2310.07063",
    archivePrefix = "arXiv",
    primaryClass = "astro-ph.HE",
    doi = "10.22323/1.444.0527",
    journal = "PoS",
    volume = "ICRC2023",
    pages = "527",
    year = "2023"
}

@article{Fuehne:2023kap,
    author = "Fuehne, Duncan and Heibges, Tobias",
    collaboration = "JEM-EUSO",
    title = "{Simulating Geomagnetic Effects on Muons in Extensive Air Showers for the EUSO-SPB2 Mission}",
    eprint = "2310.07832",
    archivePrefix = "arXiv",
    primaryClass = "astro-ph.IM",
    doi = "10.22323/1.444.0363",
    journal = "PoS",
    volume = "ICRC2023",
    pages = "363",
    year = "2023"
}

@article{Gazda:2023tbw,
    author = "Gazda, Eliza",
    collaboration = "JEM-EUSO",
    title = "{The EUSO-SPB2 Cherenkov Telescope - performance and preliminary results}",
    eprint = "2308.15628",
    archivePrefix = "arXiv",
    primaryClass = "astro-ph.IM",
    doi = "10.22323/1.444.1029",
    journal = "PoS",
    volume = "ICRC2023",
    pages = "1029",
    year = "2023"
}

@article{Matamala:2023iru,
    author = "Romero Matamala, Oscar Fernando",
    collaboration = "JEM-EUSO",
    title = "{Commissioning, Calibration, and Performance of the Cherenkov Telescope on EUSO-SPB2}",
    eprint = "2310.07561",
    archivePrefix = "arXiv",
    primaryClass = "astro-ph.HE",
    doi = "10.22323/1.444.1041",
    journal = "PoS",
    volume = "ICRC2023",
    pages = "1041",
    year = "2023"
}

@article{Diesing:2023wcq,
    author = "Diesing, Rebecca and Meyer, Stephan S. and Eser, Johannes and Bukowski, Alexa and Miller, Alex and Apfel, Jake and Beck, Gerard and Olinto, Angela V.",
    collaboration = "JEM-EUSO",
    title = "{Infrared Cloud Monitoring with UCIRC2}",
    eprint = "2310.08607",
    archivePrefix = "arXiv",
    primaryClass = "astro-ph.IM",
    doi = "10.22323/1.444.0450",
    journal = "PoS",
    volume = "ICRC2023",
    pages = "450",
    year = "2023"
}

@article{Heibges:2023yhn,
    author = "Heibges, Tobias and Posligua, Jonatan and Wistrand, Hannah and Gu\'epin, Claire and Reno, Mary Hall and Venters, Tonia M.",
    collaboration = "JEM-EUSO",
    title = "{Overview of the EUSO-SPB2 Target of Opportunity program using the Cherenkov Telescope}",
    eprint = "2310.12310",
    archivePrefix = "arXiv",
    primaryClass = "astro-ph.HE",
    doi = "10.22323/1.444.1134",
    journal = "PoS",
    volume = "ICRC2023",
    pages = "1134",
    year = "2023"
}

@article{Wistrand:2023mpb,
    author = "Wistrand, Hannah and Heibges, Tobias and Posligua, Jonatan and Gu\'epin, Claire and Reno, Mary Hall and Venters, Tonia M.",
    collaboration = "JEM-EUSO",
    title = "{The Targets of Opportunity Source Catalog for the EUSO-SPB2 Mission}",
    eprint = "2312.00920",
    archivePrefix = "arXiv",
    primaryClass = "hep-ex",
    doi = "10.22323/1.444.1185",
    journal = "PoS",
    volume = "ICRC2023",
    pages = "1185",
    year = "2023"
}

@article{Posligua:2023cdm,
    author = "Posligua, Jonatan and Heibges, Tobias and Wistrand, Hannah and Gu\'epin, Claire and Reno, Mary Hall and Venters, Tonia M.",
    collaboration = "JEM-EUSO",
    title = "{Neutrino Target-of-Opportunity Sky Coverage and Scheduler for EUSO-SPB2}",
    eprint = "2310.13827",
    archivePrefix = "arXiv",
    primaryClass = "astro-ph.HE",
    doi = "10.22323/1.444.1038",
    journal = "PoS",
    volume = "ICRC2023",
    pages = "1038",
    year = "2023"
}

@article{Garg:2022ugd,
    author = "Garg, Diksha and others",
    title = "{Neutrino propagation in the Earth and emerging charged leptons with nuPyProp}",
    eprint = "2209.15581",
    archivePrefix = "arXiv",
    primaryClass = "astro-ph.HE",
    doi = "10.1088/1475-7516/2023/01/041",
    journal = "JCAP",
    volume = "01",
    pages = "041",
    year = "2023"
}

@article{NuSpaceSim:2023ims,
    author = "Garg, Diksha and others",
    collaboration = "NuSpaceSim",
    title = "{Neutrino propagation through Earth: modeling uncertainties using nuPyProp}",
    eprint = "2308.13659",
    archivePrefix = "arXiv",
    primaryClass = "hep-ph",
    doi = "10.22323/1.444.1115",
    journal = "PoS",
    volume = "ICRC2023",
    pages = "1115",
    year = "2023"
}

@article{Guepin:2022qpl,
    author = "Gu\'epin, Claire and Kotera, Kumiko and Oikonomou, Foteini",
    title = "{High-energy neutrino transients and the future of multi-messenger astronomy}",
    eprint = "2207.12205",
    archivePrefix = "arXiv",
    primaryClass = "astro-ph.HE",
    doi = "10.1038/s42254-022-00504-9",
    journal = "Nature Rev. Phys.",
    volume = "4",
    number = "11",
    pages = "697--712",
    year = "2022"
}

@article{IceCube:2018dnn,
    author = "Aartsen, M. G. and others",
    collaboration = "IceCube, Fermi-LAT, MAGIC, AGILE, ASAS-SN, HAWC, H.E.S.S., INTEGRAL, Kanata, Kiso, Kapteyn, Liverpool Telescope, Subaru, Swift NuSTAR, VERITAS, VLA/17B-403",
    title = "{Multimessenger observations of a flaring blazar coincident with high-energy neutrino IceCube-170922A}",
    eprint = "1807.08816",
    archivePrefix = "arXiv",
    primaryClass = "astro-ph.HE",
    doi = "10.1126/science.aat1378",
    journal = "Science",
    volume = "361",
    number = "6398",
    pages = "eaat1378",
    year = "2018"
}

@article{IceCube:2018cha,
    author = "Aartsen, M. G. and others",
    collaboration = "IceCube",
    title = "{Neutrino emission from the direction of the blazar TXS 0506+056 prior to the IceCube-170922A alert}",
    eprint = "1807.08794",
    archivePrefix = "arXiv",
    primaryClass = "astro-ph.HE",
    doi = "10.1126/science.aat2890",
    journal = "Science",
    volume = "361",
    number = "6398",
    pages = "147--151",
    year = "2018"
}

@ARTICLE{2019ATel12967....1T,
       author = {{Taboada}, Ignacio and {Stein}, Robert},
        title = "{IceCube-190730A an astrophysical neutrino candidate in spatial coincidence with FSRQ PKS 1502+106}",
      journal = {The Astronomer's Telegram},
         year = 2019,
        month = jul,
       volume = {12967},
        pages = {1},
       adsurl = {https://ui.adsabs.harvard.edu/abs/2019ATel12967....1T}
}

@ARTICLE{2020A&A...640L...4G,
       author = {{Giommi}, P. and {Padovani}, P. and {Oikonomou}, F. and {Glauch}, T. and {Paiano}, S. and {Resconi}, E.},
        title = "{3HSP J095507.9+355101: A flaring extreme blazar coincident in space and time with IceCube-200107A}",
      journal = {Astron. Astrophys.},
         year = 2020,
        month = aug,
       volume = {640},
          eid = {L4},
        pages = {L4},
          doi = {10.1051/0004-6361/202038423},
archivePrefix = {arXiv},
       eprint = {2003.06405},
 primaryClass = {astro-ph.HE},
       adsurl = {https://ui.adsabs.harvard.edu/abs/2020A&A...640L...4G}
}

@article{Stein:2020xhk,
    author = "Stein, Robert and others",
    title = "{A tidal disruption event coincident with a high-energy neutrino}",
    eprint = "2005.05340",
    archivePrefix = "arXiv",
    primaryClass = "astro-ph.HE",
    doi = "10.1038/s41550-020-01295-8",
    journal = "Nature Astron.",
    volume = "5",
    number = "5",
    pages = "510--518",
    year = "2021"
}

@article{Reusch:2021ztx,
    author = "Reusch, Simeon and others",
    title = "{The candidate tidal disruption event AT2019fdr coincident with a high-energy neutrino}",
    eprint = "2111.09390",
    archivePrefix = "arXiv",
    primaryClass = "astro-ph.HE",
    journal = "arXiv",
    month = "11",
    year = "2021"
}

@article{vanVelzen:2021zsm,
    author = "van Velzen, Sjoert and others",
    title = "{Establishing accretion flares from supermassive black holes as a source of high-energy neutrinos}",
    eprint = "2111.09391",
    archivePrefix = "arXiv",
    primaryClass = "astro-ph.HE",
    doi = "10.1093/mnras/stae610",
    journal = "Mon. Not. Roy. Astron. Soc.",
    volume = "529",
    number = "3",
    pages = "2559--2576",
    year = "2024"
}

@article{IceCube:2022der,
    author = "Abbasi, R. and others",
    collaboration = "IceCube",
    title = "{Evidence for neutrino emission from the nearby active galaxy NGC 1068}",
    eprint = "2211.09972",
    archivePrefix = "arXiv",
    primaryClass = "astro-ph.HE",
    doi = "10.1126/science.abg3395",
    journal = "Science",
    volume = "378",
    number = "6619",
    pages = "538--543",
    year = "2022"
}

@article{IceCube:2023ame,
    author = "Abbasi, R. and others",
    collaboration = "IceCube",
    title = "{Observation of high-energy neutrinos from the Galactic plane}",
    eprint = "2307.04427",
    archivePrefix = "arXiv",
    primaryClass = "astro-ph.HE",
    doi = "10.1126/science.adc9818",
    journal = "Science",
    volume = "380",
    number = "6652",
    pages = "adc9818",
    year = "2023"
}

@article{LIGOScientific:2017zic,
    author = "Abbott, B. P. and others",
    collaboration = "LIGO Scientific, Virgo, Fermi-GBM, INTEGRAL",
    title = "{Gravitational Waves and Gamma-rays from a Binary Neutron Star Merger: GW170817 and GRB 170817A}",
    eprint = "1710.05834",
    archivePrefix = "arXiv",
    primaryClass = "astro-ph.HE",
    reportNumber = "LIGO-P1700308",
    doi = "10.3847/2041-8213/aa920c",
    journal = "Astrophys. J. Lett.",
    volume = "848",
    number = "2",
    pages = "L13",
    year = "2017"
}

@article{IceCube:2013cdw,
    author = "Aartsen, M. G. and others",
    collaboration = "IceCube",
    title = "{First observation of PeV-energy neutrinos with IceCube}",
    eprint = "1304.5356",
    archivePrefix = "arXiv",
    primaryClass = "astro-ph.HE",
    doi = "10.1103/PhysRevLett.111.021103",
    journal = "Phys. Rev. Lett.",
    volume = "111",
    pages = "021103",
    year = "2013"
}

@article{LSST:2008ijt,
    author = "Ivezi\'c, {\v{Z}}eljko and others",
    collaboration = "LSST",
    title = "{LSST: from Science Drivers to Reference Design and Anticipated Data Products}",
    eprint = "0805.2366",
    archivePrefix = "arXiv",
    primaryClass = "astro-ph",
    reportNumber = "SLAC-PUB-16076",
    doi = "10.3847/1538-4357/ab042c",
    journal = "Astrophys. J.",
    volume = "873",
    number = "2",
    pages = "111",
    year = "2019"
}

@inproceedings{Wei:2016eox,
    author = "Wei, J. and Cordier, B.",
    title = "{The Deep and Transient Universe in the SVOM Era: New Challenges and Opportunities - Scientific prospects of the SVOM mission}",
    eprint = "1610.06892",
    archivePrefix = "arXiv",
    primaryClass = "astro-ph.IM",
    month = "10",
    year = "2016"
}

@article{Maier:2019afm,
    author = {Maier, G. and Arrabito, L. and Bernl\"ohr, K. and Bregeon, J. and Cumani, P. and Hassan, T. and Hinton, J. and Moralejo, A.},
    collaboration = "CTA Consortium",
    title = "{Performance of the Cherenkov Telescope Array}",
    eprint = "1907.08171",
    archivePrefix = "arXiv",
    primaryClass = "astro-ph.IM",
    doi = "10.22323/1.358.0733",
    journal = "PoS",
    volume = "ICRC2019",
    pages = "733",
    year = "2020"
}

@article{Goodman:2007zz,
    author = "Goodman, Jordan A.",
    editor = "Valdez, Heriberto Castilla and Perez, Miguel A. and D'Olivo, Juan Carlos",
    collaboration = "HAWC",
    title = "{HAWC: A next generation all-sky gamma-ray telescope}",
    doi = "10.1063/1.2751958",
    journal = "AIP Conf. Proc.",
    volume = "917",
    number = "1",
    pages = "206--209",
    year = "2007"
}

@article{LHAASO:2019qtb,
    author = "Addazi, Andrea and others",
    collaboration = "LHAASO",
    title = "{The Large High Altitude Air Shower Observatory (LHAASO) Science Book (2021 Edition)}",
    eprint = "1905.02773",
    archivePrefix = "arXiv",
    primaryClass = "astro-ph.HE",
    journal = "Chin. Phys. C",
    volume = "46",
    pages = "035001--035007",
    year = "2022"
}

@article{Krizmanic:2023pwf,
    author = "Krizmanic, John",
    collaboration = "NuSpaceSim",
    title = "{$\nu$SpaceSim: A Comprehensive Simulation Package to Model the Optical and Radio Signals from Extensive Air Showers Induced by Cosmic Neutrinos and Measured by Space-based Experiments}",
    doi = "10.22323/1.444.1110",
    journal = "PoS",
    volume = "ICRC2023",
    pages = "1110",
    year = "2023"
}

@INPROCEEDINGS{2008ICRC....3.1341W,
       author = {{Wakely}, S.~P. and {Horan}, D.},
        title = "{TeVCat: An online catalog for Very High Energy Gamma-Ray Astronomy}",
    booktitle = {International Cosmic Ray Conference},
         year = 2008,
       series = {International Cosmic Ray Conference},
       volume = {3},
        month = jan,
        pages = {1341-1344},
       adsurl = {https://ui.adsabs.harvard.edu/abs/2008ICRC....3.1341W}
}

@article{Adams:2022oko,
    author = "Adams, J. H. and others",
    title = "{A Review of the EUSO-Balloon Pathfinder for the JEM-EUSO Program}",
    doi = "10.1007/s11214-022-00870-x",
    journal = "Space Sci. Rev.",
    volume = "218",
    number = "1",
    pages = "3",
    year = "2022"
}

@article{GCN,
title="GCN: The Gamma-ray Coordinates Network (TAN: Transient Astronomy Network)",
howpublished={\url{https://gcn.gsfc.nasa.gov/}}
}

@article{TNS,
title="Transient Name Server",
howpublished={\url{https://www.wis-tns.org/}}
}

@article{Gill:2024mqb,
    author = "Gill, Ajay S. and others",
    title = "{SuperBIT Superpressure Flight Instrument Overview and Performance: Near diffraction-limited Astronomical Imaging from the Stratosphere}",
    eprint = "2408.01847",
    archivePrefix = "arXiv",
    primaryClass = "astro-ph.IM",
    doi = "10.3847/1538-3881/ad5840",
    journal = "Ap. J.",
    volume = "168",
    pages = "85",
    year = "2024"
}

@article{Battisti:2024jjy,
    author = "Battisti, Matteo and Eser, Johannes and Olinto, Angela and Osteria, Giuseppe",
    collaboration = "JEM-EUSO",
    title = "{POEMMA-Balloon with Radio: A balloon-born multi-messenger multi-detector observatory}",
    eprint = "2409.06753",
    archivePrefix = "arXiv",
    primaryClass = "astro-ph.IM",
    doi = "10.1016/j.nima.2024.169819",
    journal = "Nucl. Instrum. Meth. A",
    volume = "1069",
    pages = "169819",
    year = "2024"
}

@ARTICLE{Zotti_Hoffmann_Wolf_Chereau_Chereau_2021,
       author = {{Zotti}, Georg and {Hoffmann}, Susanne M and {Wolf}, Alexander and {Ch{\'e}reau}, Fabien and {Ch{\'e}reau}, Guillaume},
        title = "{The Simulated Sky: Stellarium for Cultural Astronomy Research}",
      journal = {arXiv e-prints},
         year = 2021,
        month = mar,
          eid = {arXiv:2104.01019},
        pages = {arXiv:2104.01019},
          doi = {10.48550/arXiv.2104.01019},
archivePrefix = {arXiv},
       eprint = {2104.01019},
 primaryClass = {astro-ph.IM},
       adsurl = {https://ui.adsabs.harvard.edu/abs/2021arXiv210401019Z}
}

@article{Zotti_Wolf_2022,
    title={Stellarium: Finally at Version 1.0! And Beyond}, 
    volume={8},
    url={https://journal.equinoxpub.com/JSA/article/view/25608}, 
    DOI={10.1558/jsa.25608},
    abstractNote={
    Stellarium: Finally at Version 1.0! And Beyond
    },
    number={2},
    journal={Journal of Skyscape Archaeology},
    author={Zotti, Georg and Wolf, Alexander},
    year={2023},
    month={Feb.},
    pages={332–334} 
}

@article{KM3NeT:2025npi,
    author = "Aiello, S. and others",
    collaboration = "KM3NeT",
    title = "{Observation of an ultra-high-energy cosmic neutrino with KM3NeT}",
    doi = "10.1038/s41586-024-08543-1",
    journal = "Nature",
    volume = "638",
    number = "8050",
    pages = "376--382",
    year = "2025",
    note = "[Erratum: Nature 640, E3 (2025)]"
}

@article{POEMMA:2020ykm,
    author = "Olinto, A. V. and others",
    collaboration = "POEMMA",
    title = "{The POEMMA (Probe of Extreme Multi-Messenger Astrophysics) observatory}",
    eprint = "2012.07945",
    archivePrefix = "arXiv",
    primaryClass = "astro-ph.IM",
    doi = "10.1088/1475-7516/2021/06/007",
    journal = "JCAP",
    volume = "06",
    pages = "007",
    year = "2021"
}

@article{GRAND:2018iaj,
    author = "\'Alvarez-Mu\~niz, Jaime and others",
    collaboration = "GRAND",
    title = "{The Giant Radio Array for Neutrino Detection (GRAND): Science and Design}",
    eprint = "1810.09994",
    archivePrefix = "arXiv",
    primaryClass = "astro-ph.HE",
    doi = "10.1007/s11433-018-9385-7",
    journal = "Sci. China Phys. Mech. Astron.",
    volume = "63",
    number = "1",
    pages = "219501",
    year = "2020"
}

@article{Wissel:2020sec,
    author = "Wissel, Stephanie and others",
    title = "{Prospects for high-elevation radio detection of \ensuremath{>}100 PeV tau neutrinos}",
    eprint = "2004.12718",
    archivePrefix = "arXiv",
    primaryClass = "astro-ph.IM",
    doi = "10.1088/1475-7516/2020/11/065",
    journal = "JCAP",
    volume = "11",
    pages = "065",
    year = "2020"
}

@ARTICLE{2022TNSTR1521....1T,
       author = {{Tonry}, J. and {Denneau}, L. and {Weiland}, H. and {Erasmus}, N. and {Koorts}, W. and {Anderson}, J. and {Jordan}, A. and {Suc}, V. and {Smith}, K.~W. and {Srivastav}, S. and {Young}, D.~R. and {Smartt}, S.~J. and {Gillanders}, J. and {Fulton}, M. and {McCollum}, M. and {Moore}, T. and {Shingles}, L. and {Rest}, A. and {Chen}, T.~W. and {Nicholl}, M. and {Pacheco}, D. and {Stubbs}, C.},
        title = "{ATLAS Transient Discovery Report for 2022-06-02}",
      journal = {Transient Name Server Discovery Report},
         year = 2022,
        month = jun,
       volume = {2022-1521},
        pages = {1},
       adsurl = {https://ui.adsabs.harvard.edu/abs/2022TNSTR1521....1T}
}

@article{Reichherzer:2021pfe,
    author = {Reichherzer, P. and Sch{\"u}ssler, F. and Lefranc, V. and Yusafzai, A. and Alkan, A. K. and Ashkar, H. and Becker Tjus, J.},
    title = "{Astro-COLIBRI{\textemdash}The COincidence LIBrary for Real-time Inquiry for Multimessenger Astrophysics}",
    eprint = "2109.01672",
    archivePrefix = "arXiv",
    primaryClass = "astro-ph.IM",
    doi = "10.3847/1538-4365/ac1517",
    journal = "Astrophys. J. Supp.",
    volume = "256",
    number = "1",
    pages = "5",
    year = "2021"
}

@article{Venters:2019xwi,
    author = "Venters, Tonia M. and Reno, Mary Hall and Krizmanic, John F. and Anchordoqui, Luis A. and Gu\'epin, Claire and Olinto, Angela V.",
    title = "{POEMMA's Target of Opportunity Sensitivity to Cosmic Neutrino Transient Sources}",
    eprint = "1906.07209",
    archivePrefix = "arXiv",
    primaryClass = "astro-ph.HE",
    doi = "10.1103/PhysRevD.102.123013",
    journal = "Phys. Rev. D",
    volume = "102",
    pages = "123013",
    year = "2020"
}

@article{Anchordoqui:2025xug,
    author = "Anchordoqui, Luis A. and Halzen, Francis and Lust, Dieter",
    title = "{Neutrinos from Primordial Black Holes in Theories with Extra Dimensions}",
    eprint = "2505.23414",
    archivePrefix = "arXiv",
    primaryClass = "hep-ph",
    reportNumber = "MPP-2025-110; LMU-ASC 14/25",
    month = "5",
    year = "2025"
}

@article{Dev:2025czz,
    author = "Dev, P. S. Bhupal and Dutta, Bhaskar and Karthikeyan, Aparajitha and Maitra, Writasree and Strigari, Louis E. and Verma, Ankur",
    title = "{`Dark' Matter Effect as a Novel Solution to the KM3-230213A Puzzle}",
    eprint = "2505.22754",
    archivePrefix = "arXiv",
    primaryClass = "hep-ph",
    reportNumber = "MI-HET-859",
    month = "5",
    year = "2025"
}

@article{Farzan:2025ydi,
    author = "Farzan, Yasaman and Hostert, Matheus",
    title = "{Astrophysical sources of dark particles as a solution to the KM3NeT and IceCube tension over KM3-230213A}",
    eprint = "2505.22711",
    archivePrefix = "arXiv",
    primaryClass = "hep-ph",
    month = "5",
    year = "2025"
}

@article{Murase:2025uwv,
    author = "Murase, Kohta and Narita, Yuma and Yin, Wen",
    title = "{Superheavy dark matter from the natural inflation in light of the highest-energy astroparticle events}",
    eprint = "2504.15272",
    archivePrefix = "arXiv",
    primaryClass = "hep-ph",
    reportNumber = "TU-1260",
    month = "4",
    year = "2025"
}

@article{Brdar:2025azm,
    author = "Brdar, Vedran and Chattopadhyay, Dibya S.",
    title = "{Does the 220 PeV Event at KM3NeT Point to New Physics?}",
    eprint = "2502.21299",
    archivePrefix = "arXiv",
    primaryClass = "hep-ph",
    month = "2",
    year = "2025"
}

@article{Neronov:2025jfj,
    author = "Neronov, Andrii and Oikonomou, Foteini and Semikoz, Dmitri",
    title = "{KM3-230213A: An Ultra-High Energy Neutrino from a Year-Long Astrophysical Transient}",
    eprint = "2502.12986",
    archivePrefix = "arXiv",
    primaryClass = "astro-ph.HE",
    month = "2",
    year = "2025"
}

@article{Fang:2025nzg,
    author = "Fang, Ke and Halzen, Francis and Hooper, Dan",
    title = "{Cascaded Gamma-Ray Emission Associated with the KM3NeT Ultrahigh-energy Event KM3-230213A}",
    eprint = "2502.09545",
    archivePrefix = "arXiv",
    primaryClass = "astro-ph.HE",
    doi = "10.3847/2041-8213/adbbec",
    journal = "Astrophys. J. Lett.",
    volume = "982",
    number = "1",
    pages = "L16",
    year = "2025"
}

@article{Adams:2025owi,
    author = "Adams, Jr., J. H. and others",
    title = "{The Extreme Universe Observatory on a Super-Pressure Balloon II: Mission, Payload, and Flight}",
    eprint = "2505.20762",
    archivePrefix = "arXiv",
    primaryClass = "astro-ph.HE",
    month = "5",
    year = "2025"
}

@article{Plebaniak:2025ayq,
    author = "Plebaniak, Zbigniew Dariusz",
    collaboration = "JEM-EUSO",
    title = "{From Ground to Space: An Overview of the JEM-EUSO Program for the Study of UHECRs and Astrophysical Neutrinos}",
    eprint = "2511.17139",
    archivePrefix = "arXiv",
    primaryClass = "astro-ph.IM",
    doi = "10.22323/1.501.0360",
    journal = "PoS",
    volume = "ICRC2025",
    pages = "360",
    year = "2025"
}

@article{Eser:2025fcw,
    author = "Eser, Johannes and Olinto, Angela V. and Osteria, Giuseppe",
    title = "{POEMMA-Balloon with Radio: An Overview}",
    eprint = "2509.04302",
    archivePrefix = "arXiv",
    primaryClass = "astro-ph.IM",
    doi = "10.22323/1.501.0249",
    journal = "PoS",
    volume = "ICRC2025",
    pages = "249",
    year = "2025"
}

@article{Reno:2025tpw,
    author = "Reno, Mary Hall and Krizmanic, John F.",
    title = "{NuSpaceSim: A Comprehensive Simulation Package for Modeling the Measurement of Cosmic Neutrinos using the Earth as the Neutrino Target and Space-based Detectors}",
    eprint = "2509.15469",
    archivePrefix = "arXiv",
    primaryClass = "astro-ph.HE",
    doi = "10.22323/1.501.1082",
    journal = "PoS",
    volume = "ICRC2025",
    pages = "1082",
    year = "2025"
}

@article{Heibges:2025yvw,
    author = "Heibges, Tobias and Gu{\'e}pin-Detrigne, Claire and Garg, Diksha and Kupari, Luke and Reno, Mary Hall and Venters, Tonia M. and Wiencke, Lawrence",
    title = "{Using the Cherenkov Telescope onboard EUSO-SPB2 for Target of Opportunity searches of very high energy neutrino sources}",
    eprint = "2511.10332",
    archivePrefix = "arXiv",
    primaryClass = "astro-ph.HE",
    doi = "10.22323/1.501.1051",
    journal = "PoS",
    volume = "ICRC2025",
    pages = "1051",
    year = "2025"
}

@article{GRAND:2025rps,
    author = "Kotera, Kumiko and others",
    collaboration = "GRAND, Beacon",
    title = "{The Hybrid Elevated Radio Observatory for Neutrinos (HERON) Project}",
    eprint = "2507.04382",
    archivePrefix = "arXiv",
    primaryClass = "astro-ph.IM",
    reportNumber = "PoS(ICRC2025)1078",
    doi = "10.22323/1.501.1078",
    journal = "PoS",
    volume = "ICRC2025",
    pages = "1078",
    year = "2025"
}

@ARTICLE{2018AJ....155..128M,
       author = {{Morris}, Brett M. and {Tollerud}, Erik and {Sip{\H{o}}cz}, Brigitta and {Deil}, Christoph and {Douglas}, Stephanie T. and {Berlanga Medina}, Jazmin and {Vyhmeister}, Karl and {Smith}, Toby R. and {Littlefair}, Stuart and {Price-Whelan}, Adrian M. and {Gee}, Wilfred T. and {Jeschke}, Eric},
        title = "{astroplan: An Open Source Observation Planning Package in Python}",
      journal = {\aj},
         year = 2018,
        month = mar,
       volume = {155},
       number = {3},
          eid = {128},
        pages = {128},
          doi = {10.3847/1538-3881/aaa47e},
archivePrefix = {arXiv},
       eprint = {1712.09631},
 primaryClass = {astro-ph.IM},
       adsurl = {https://ui.adsabs.harvard.edu/abs/2018AJ....155..128M}
}

@INPROCEEDINGS{2006smci.conf...80F,
       author = {{Frank}, J.},
        title = "{SOFIA's challenge: automated scheduling of airborne astronomy observations}",
    booktitle = {2nd IEEE International Conference on Space Mission Challenges for Information Technology (SMC-IT'06)},
         year = 2006,
        month = jan,
    publisher = {IEEE},
          eid = {80},
        pages = {80},
          doi = {10.1109/SMC-IT.2006.70},
       adsurl = {https://ui.adsabs.harvard.edu/abs/2006smci.conf...80F}
}

@article{Adams:2026cpe,
    author = "Adams, J. and others",
    title = "{POEMMA-Balloon with Radio: A multi-messenger, multi-detector balloon payload}",
    eprint = "2601.19997",
    archivePrefix = "arXiv",
    primaryClass = "astro-ph.IM",
    month = "1",
    year = "2026"
}
}

\end{document}